\documentclass[pdflatex,sn-mathphys]{sn-jnl}
\usepackage{graphicx}
\usepackage{hyperref}
\usepackage{pifont}
\usepackage{threeparttable}
\usepackage{makecell}
\usepackage{multicol}
\usepackage{multirow}
\usepackage{booktabs}
\usepackage{makecell}
\usepackage{adjustbox}
\usepackage{bm}

\jyear{2023}%

\theoremstyle{thmstyleone}%
\theoremstyle{thmstyletwo}%

\theoremstyle{thmstylethree}%

\begin{document}

\title[DeepRelax]{Scalable Crystal Structure Relaxation Using an Iteration-Free Deep Generative Model with Uncertainty Quantification}

\author[1,2]{\fnm{Ziduo} \sur{Yang}}
\equalcont

\author[1]{\fnm{Yi-Ming} \sur{Zhao}}
\equalcont{These authors contributed equally to this work.}

\author[3]{\fnm{Xian} \sur{Wang}}

\author[1]{\fnm{Xiaoqing} \sur{Liu}}

\author[1]{\fnm{Xiuying} \sur{Zhang}}

\author[1]{\fnm{Yifan} \sur{Li}}

\author[1,2]{\fnm{Qiujie} \sur{Lv}}

\author*[4,5,6,7,8]{\fnm{Calvin Yu-Chian} \sur{Chen}}\email{cy@pku.edu.cn}

\author*[1,9]{\fnm{Lei} \sur{Shen}}\email{shenlei@nus.edu.sg}

\affil[1]{\orgdiv{Department of Mechanical Engineering}, \orgname{National University of Singapore}, \orgaddress{\city{Singapore}, \postcode{117575}, \country{Singapore}}}

\affil[2]{\orgdiv{Artificial Intelligence Medical Research Center, School of Intelligent Systems Engineering}, \orgname{Shenzhen Campus of Sun Yat-sen University}, \orgaddress{\city{Shenzhen}, \postcode{518107}, \country{China}}}

\affil[3]{\orgdiv{Department of Physics}, \orgname{National University of Singapore}, \orgaddress{\city{Singapore}, \postcode{117551}, \country{Singapore}}}

\affil[4]{\orgdiv{AI for Science (AI4S)-Preferred Program, School of Electronic and Computer Engineering}, \orgname{Peking University Shenzhen Graduate School}, \orgaddress{\city{Shenzhen}, \postcode{518055}, \country{China}}}

\affil[5]{\orgdiv{State Key Laboratory of Chemical Oncogenomics, School of Chemical Biology and Biotechnology}, \orgname{Peking University Shenzhen Graduate School}, \orgaddress{\city{Shenzhen}, \postcode{518055}, \country{China}}}

\affil[6]{\orgdiv{Department of Medical Research}, \orgname{China Medical University Hospital}, \orgaddress{\city{Taichung}, \postcode{40447}, \country{Taiwan}}}

\affil[7]{\orgdiv{Department of Bioinformatics and Medical Engineering}, \orgname{Asia University}, \orgaddress{\city{Taichung}, \postcode{41354}, \country{Taiwan}}}

\affil[8]{\orgdiv{Guangdong L-Med Biotechnology Co., Ltd}, \orgname{Meizhou}, \orgaddress{\city{Guangdong}, \postcode{514699}, \country{China}}}

\affil[9]{\orgname{National University of Singapore (Chongqing) Research Institute}, \orgaddress{\city{Chongqing}, \postcode{401123}, \country{China}}}


\abstract{In computational molecular and materials science, determining equilibrium structures is the crucial first step for accurate subsequent property calculations. However, the recent discovery of millions of new crystals and complex twisted structures has challenged traditional computational methods, both ab initio and machine-learning-based, due to their computationally intensive iterative processes. To address these scalability issues, here we introduce DeepRelax, a deep generative model capable of performing geometric crystal structure relaxation rapidly and without iterations. DeepRelax learns the equilibrium structural distribution, enabling it to predict relaxed structures directly from their unrelaxed ones. The ability to perform structural relaxation at the millisecond level per structure, combined with the scalability of parallel processing, makes DeepRelax particularly useful for large-scale virtual screening. We demonstrate DeepRelax's reliability and robustness by applying it to five diverse databases, including oxides, Materials Project, two-dimensional materials, van der Waals crystals, and crystals with point defects. DeepRelax consistently shows high accuracy and efficiency, validated by density functional theory calculations. Finally, we enhance its trustworthiness by integrating uncertainty quantification. This work significantly accelerates computational workflows, offering a robust and trustworthy machine-learning method for material discovery and advancing the application of AI for science. Code for DeepRelax is available at \url{https://github.com/Shen-Group/DeepRelax}.}




\maketitle

\section{Introduction}
\begin{figure}[!htb]
  \centering
  \includegraphics[width=11.8cm]{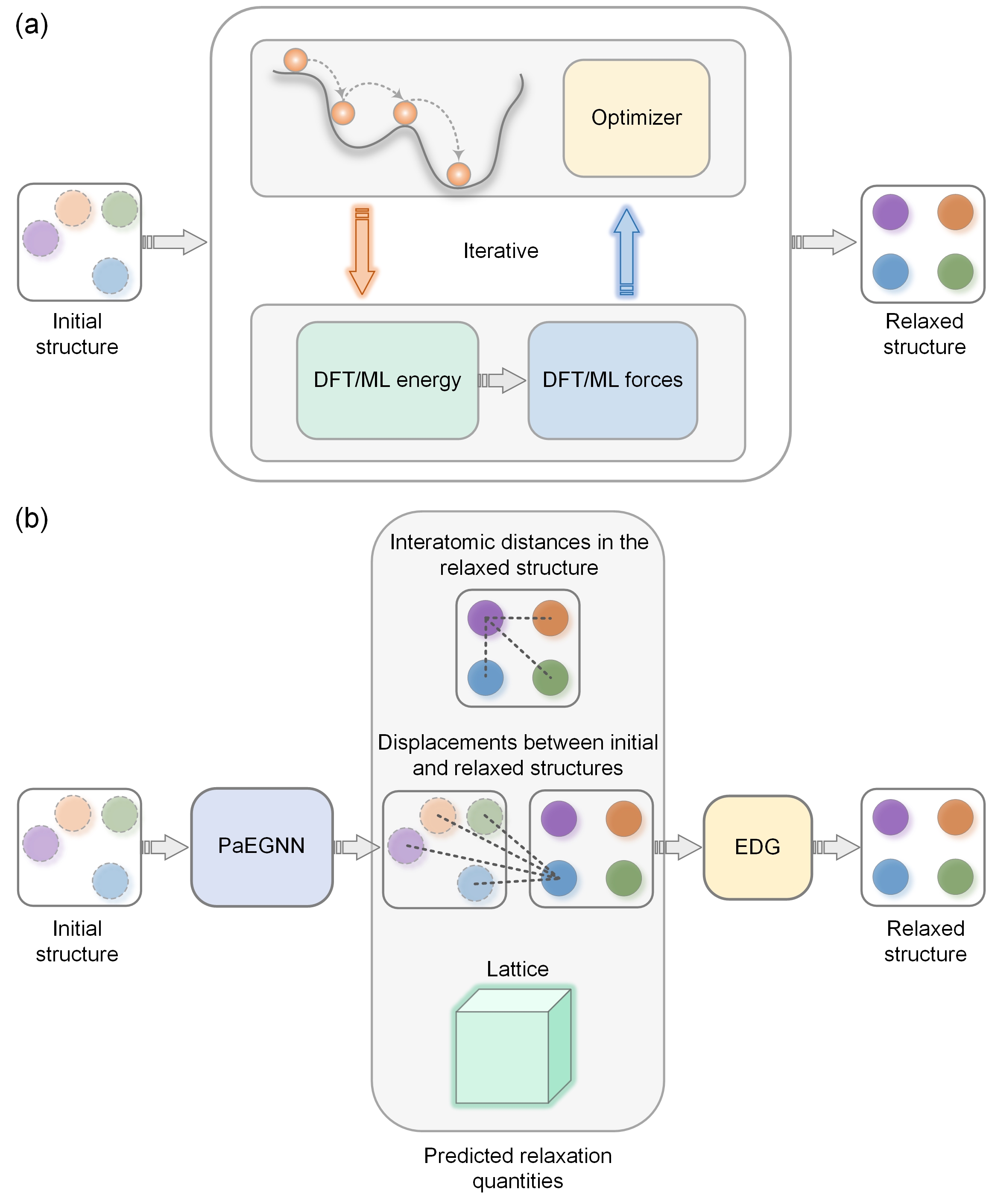}
  \caption{An overview of ML methods for crystal structure relaxation. (a) Iterative ML methods that iteratively estimate energy and force to determine the equilibrium structure. (b) Illustration of our proposed DeepRelax method, which employs a periodicity-aware equivariant graph neural network (PaEGNN) to directly predict the relaxation quantities. Euclidean distance geometry (EDG) is then used to determine the final relaxed structure that satisfies the predicted relaxation quantities.}
  \label{fgr:overview}
\end{figure}

Atomic structural relaxation is usually the first step and foundation for further computational analysis of properties in computational chemistry, physics, materials science, and medicine. This includes applications such as chemical reactions on surfaces, complex defects in semiconductor heterostructures, and drug design. To date, computational relaxation algorithms have typically been achieved using iterative optimization, such as traditional ab initio methods as shown in Fig. \ref{fgr:overview}(a). For example, each iteration in density functional theory (DFT) calculations involves solving the Schrödinger equation to determine the electronic density distribution, from which the total energy of the system can be calculated. The forces on each atom, derived from differentiating this energy with respect to atomic positions, guide atomic movements to lower the system's energy, typically using optimization algorithms. Despite its effectiveness, the high computational demands and poor scalability of DFT limit its applications across high-dimensional chemical and structural spaces \cite{zuo2021accelerating}, such as the complex chemical reaction surfaces, doped semiconductor interfaces, or in the structural relaxation of the 2.2 million new crystals recently identified by DeepMind \cite{merchant2023scaling}. It is worth noting that the discovery of huge new materials has been significantly accelerated by high-throughput DFT calculations \cite{ong2013python, saal2013materials, curtarolo2012aflow, zhou20192dmatpedia} and advanced machine learning (ML) algorithms \cite{chen2021phase, xie2021crystal, zeni2023mattergen, zhao2023physics}, which is promoting the development of more efficient relaxation algorithms.  

ML has emerged as a promising alternative for predicting relaxed structures \cite{chen2022universal, deng2023chgnet, zuo2021accelerating, mosquera2024machine, kolluru2022open, kim2023structure, yoon2020differentiable, wang2024concurrent, Omee2024}. As conventional iterative optimization, iterative ML approaches \cite{chen2022universal, deng2023chgnet, zuo2021accelerating, kolluru2022open, mosquera2024machine, wang2024concurrent, Omee2024} utilize surrogate ML models to approximate energy and forces at each iteration, as shown in Fig. \ref{fgr:overview}(a), thereby circumventing the need to solve the computationally intensive Schrödinger equation. A typical example is the defect engineering in crystalline materials \cite{kazeev2023sparse, mosquera2023identifying, huang2023unveiling}. Mosquera-Lois et al. \cite{mosquera2024machine} and Jiang et al. \cite{jiang2024machine} demonstrated that ML surrogate models could accelerate the optimization of crystals with defects. These ML models can retain DFT-level accuracy by training on extensive databases containing detailed information on structural relaxations, including energy, forces, and stress. 

However, there are two primary challenges in current iterative ML structural optimizers: training data limitations and non-scalability. Their training dataset must include full or partial intermediate steps of DFT relaxation. However, almost all publicized material databases, such as ICSD \cite{belsky2002new} and 2DMatPedia \cite{ zhou20192dmatpedia}, do not provide such structural information, potentially limiting the application of iterative ML methods. The other challenge is that the large-scale parallel processing capability of iterative ML methods is limited due to their step-by-step nature. To address this, Yoon et al. \cite{yoon2020differentiable} developed a model called DOGSS, and Kim et al. \cite{kim2023structure} proposed a model named Cryslator. Both conceptually introduce direct ML approaches to predict the final relaxed structures from their unrelaxed counterparts. However, these approaches have only been validated on specific datasets or systems, and their universal applicability to diverse datasets or systems remains unproven.

In this work, we introduce DeepRelax, a scalable, universal, and trustworthy deep generative model designed for direct structural relaxation. DeepRelax requires only the initial crystal structures to predict equilibrium structures in just a few hundred milliseconds on a single GPU. Furthermore, DeepRelax can efficiently handle multiple crystal structures in parallel by organizing them into mini-batches for simultaneous processing. This capability is especially advantageous in large-scale virtual screening, where rapid assessment of numerous unknown crystal configurations is essential. To demonstrate the reliability and robustness, we evaluate DeepRelax across five different datasets, including diverse 3D and 2D materials: the Materials Project (MP) \cite{jain2013commentary}, X-Mn-O oxides \cite{kim2020generative, kim2023structure}, the Computational 2D Materials Database (C2DB) \cite{haastrup2018computational, gjerding2021recent, lyngby2022data}, layered van der Waals crystals, and 2D structures with point defects \cite{huang2023unveiling, kazeev2023sparse}. DeepRelax not only demonstrates superior performance compared to other direct ML methods but also exhibits competitive accuracy to the leading iterative ML model, M3GNet \cite{chen2022universal}, while being approximately 100 times faster in terms of speed. Moreover, we conduct DFT calculations to assess the energy of DeepRelax's predicted structures, confirming our model's ability to identify energetically favorable configurations. Additionally, DeepRelax employs an uncertainty quantification method to assess the trustworthiness of the model. Finally, we would like to highlight that the aim of using DeepRelax is not to replace DFT relaxation, but to make the predicted structures close to the DFT-relaxed configuration. Thus, the DFT method can rapidly complete the relaxation with only a few steps, significantly speeding up the traditional ab initio relaxation process, especially for complex structures. 

\section{Results}
\subsection{DeepRelax architecture}
DeepRelax emerges as a solution to the computational bottlenecks faced in DFT methods for crystal structure relaxation. Fig. \ref{fgr:overview}(b) shows the workflow of DeepRelax, which takes an unrelaxed crystal structure as input and uses a periodicity-aware equivariant graph neural network (PaEGNN) to predict the relaxation quantities, including interatomic distances in the relaxed structure, displacements between the initial and relaxed structures, and the lattice matrix of the relaxed structure. DeepRelax then employs a numerical Euclidean distance geometry (EDG) solver to determine the relaxed structure that satisfies the predicted relaxation quantities. In addition, DeepRelax also quantifies bond-level uncertainty for each predicted interatomic distance and displacement. Aggregating these bond-level uncertainties allows for the computation of the system-level uncertainty, offering valuable insights into the trustworthiness of the model.

A notable feature of PaEGNN, distinguishing it from previous graph neural networks (GNNs) \cite{schutt2021equivariant, xie2018crystal}, is the explicit differentiation of atoms in various translated cells to encode periodic boundary conditions (PBCs) using a unit cell offset encoding (UCOE). Additionally, its design ensures equivariance, facilitating active exploration of crystal symmetries and thus providing a richer geometric representation of crystal structures.  

\subsection{Benchmark on X-Mn-O dataset}
For our initial benchmarking, we utilize the X-Mn-O dataset, a hypothetical elemental substitution database previously employed for photoanode application studies \cite{noh2019unveiling, kim2020generative}. This dataset derives from the MP database, featuring prototype ternary structures that undergo elemental substitution with X elements (Mg, Ca, Ba, and Sr). It consists of 28,579 data pairs, with each comprising an unrelaxed structure and its corresponding DFT-relaxed state. The dataset is divided into training ($N=22,863$), validation ($N=2,857$), and test ($N=2,859$) sets, adhering to an 8:1:1 ratio. As illustrated in Suppl. Fig. 1, there are significant structural differences between the unrelaxed and DFT-relaxed structures within this dataset.

We conduct a comparative analysis of DeepRelax against the state-of-the-art (SOTA) benchmark model, Cryslator \cite{kim2023structure}. Additionally, we incorporate two types of equivariant graph neural networks—PAINN \cite{schutt2021equivariant} and EGNN \cite{satorras2021n}—into our analysis (see Subsection \ref{sec:implementation} for the details). The choice of equivariant models is informed by recent reports highlighting their accuracy in direct coordinate prediction for structural analysis \cite{satorras2021n, zhang2023efficient, dong2023equivariant}. To ensure a fair comparison, we use the same training, validation, and testing sets across all models. As a baseline measure, we introduce a Dummy model, which simply returns the input initial structure as its output. This serves as a control reference in our evaluation process.

To evaluate model performance, we use the mean absolute error (MAE) of Cartesian coordinates, bond lengths, lattice matrix, and cell volume to measure the consistency between predicted and DFT-relaxed structures. Additionally, we calculate the match rate—a measure of how closely predicted relaxed structures align with their ground truth counterparts within a defined tolerance, as determined by Pymatgen \cite{ong2013python}. Detailed descriptions of these metrics are provided in Subsection \ref{sec:performance_indicators}.

\begin{figure}[htb]
  \centering
  \includegraphics[width=11.8cm]{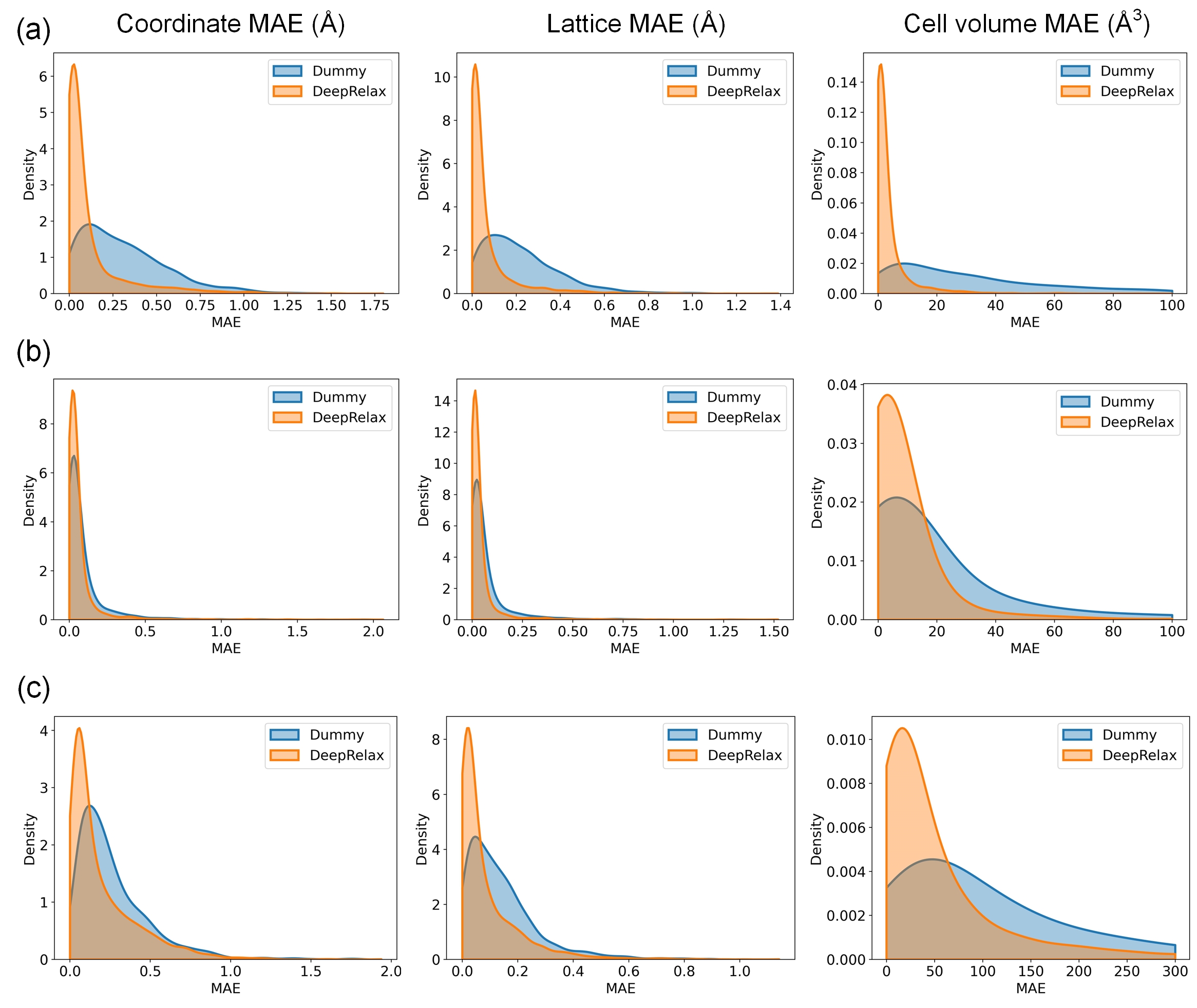}
  \caption{Distribution of MAE for predicted structures by the Dummy model and DeepRelax. (a) X-Mn-O dataset, (b) MP dataset, and (c) C2DB dataset. Each subfigure, from left to right, displays the MAE for coordinates ($\rm \AA$), lattice matrices ($\rm \AA$), and cell volumes ($\rm \AA^3$), respectively. Source data are provided with this paper.}
  \label{fgr:pred_distribution}
\end{figure}

\begin{figure}[htb]
  \centering
  \includegraphics[width=11.8cm]{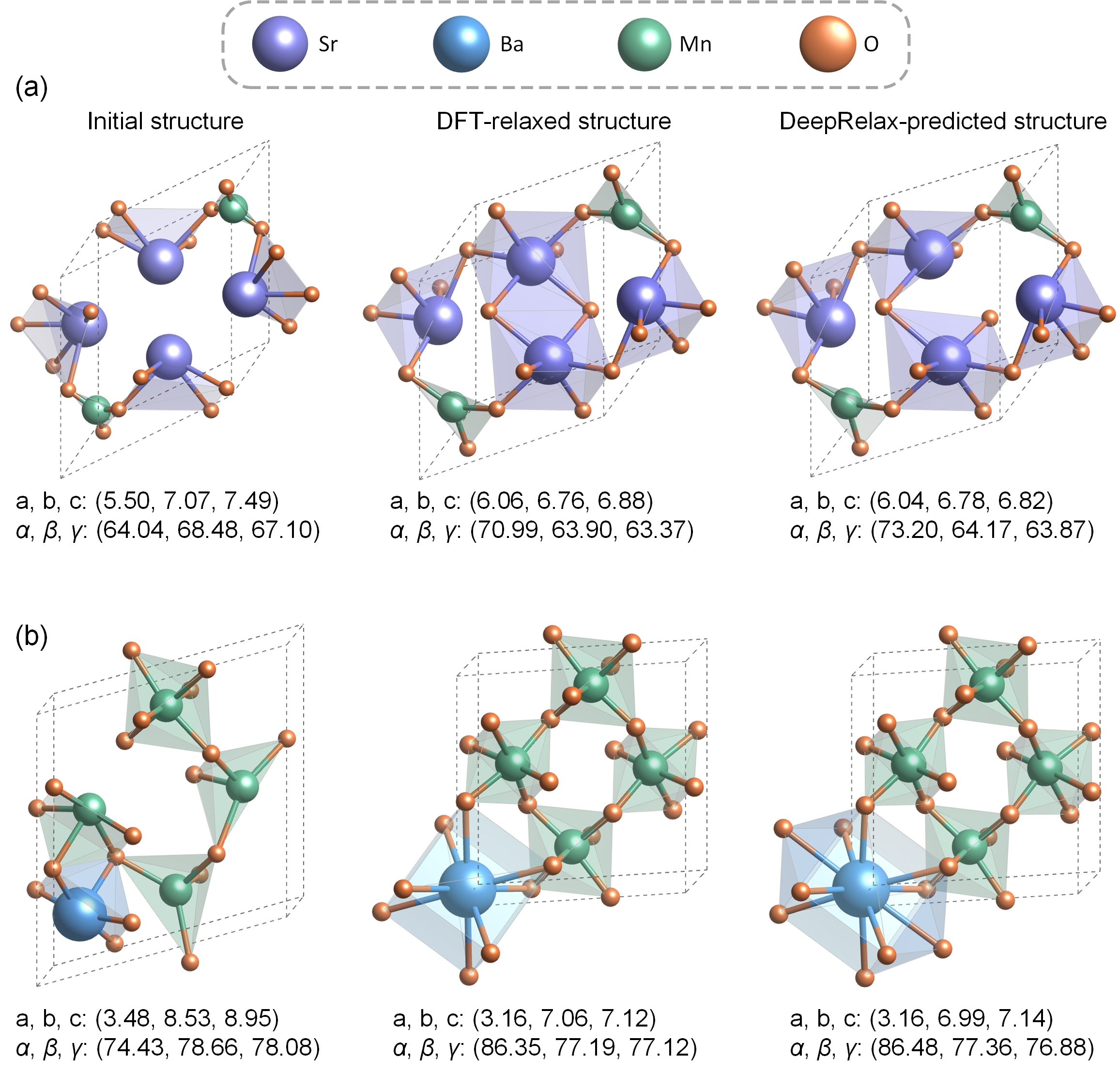}
  \caption{Visualization of two crystal structures relaxed by DeepRelax. (a) $\rm Sr_4Mn_2O_6$ and (b) $\rm Ba_1Mn_4O_8$, where \(a\), \(b\), and \(c\) are lattice constants in angstroms (Å), and \(\alpha\), \(\beta\), and \(\gamma\) are angles in degrees ($^\circ$). The results demonstrate close agreement between DeepRelax-predicted structures and DFT-relaxed structures.}
  \label{fgr:visualization}
\end{figure}

Table \ref{tbl:X-Mn-O} presents the comparative results, showing that DeepRelax greatly outperforms other baselines. Notably, DeepRelax shows a remarkable improvement in prediction accuracy over the Dummy model, with enhancements of 63.06\%, 68.30\%, 71.49\%, 89.63\%, and 30.71\% across coordinates, bond lengths, lattice, cell volumes, and match rate, respectively. Moreover, DeepRelax surpasses the previous SOTA model, Cryslator, by 8.66\% in coordinate prediction, and 45.16\% in cell volume estimation. Fig. \ref{fgr:pred_distribution}(a) shows the distribution of MAE for coordinates, lattice matrix, and cell volumes as predicted by the Dummy model and DeepRelax. DeepRelax demonstrates a notable leftward skewness in its distribution, signifying a tendency to predict structures that closely approach the DFT-relaxed state. To visualize the performance of DeepRelax, we take two typical structures, $\rm Sr_4Mn_2O_6$ and $\rm Ba_1Mn_4O_8$ from the X-Mn-O database (see Fig. \ref{fgr:visualization}), and relax them using DeepRelax. As can be seen, the DeepRelax-predicted structures are highly consistent with the DFT-relaxed ones. The results demonstrate close agreement with DFT-relaxed structures. More DFT validations are in Subsection \ref{sec:DFT_validations}.

\begin{table}[htb]
\centering
\caption{Comparative results of DeepRelax and baseline models on the X-Mn-O dataset, evaluated based on MAE of coordinates ($\rm \AA$), bond length ($\rm \AA$), lattice ($\rm \AA$), cell volume ($\rm \AA^3$), and match rate (\%) between the predicted and DFT-relaxed structures}
\label{tbl:X-Mn-O}
\tabcolsep=0.25cm
\begin{tabular}{llllll}
\toprule
Model     & Coordinates & Bond length & Lattice & Cell volume & Match rate \\ \midrule
Dummy     & 0.314           & 0.429           & 0.221    & 32.8           & 64.8            \\
PAINN     & 0.159           & 0.175           & 0.066    & 3.8          & 81.2            \\
EGNN      & 0.166           & 0.189           & 0.066    & 4.2          & 77.5            \\
Cryslator$^*$ & 0.127           & -               & -        & 6.2              & 83.7            \\
DeepRelax & \textbf{0.116}  & \textbf{0.136}  & \textbf{0.063}    & \textbf{3.4}   & \textbf{84.7} \\ \bottomrule
\end{tabular}
\\
\begin{flushleft}
\footnotesize{The best performance in each metric is highlighted in bold.
\\
$^*$The results of Cryslator are taken from \cite{kim2023structure}. DeepRelax is evaluated on the same training, validation, and testing sets as Cryslator for a fair comparison.}
\end{flushleft}
\end{table}

\subsection{Benchmark on Materials Project}
To demonstrate DeepRelax's universal applicability across various elements of the periodic table and diverse crystal types, we conduct further evaluations using the Materials Project dataset \cite{chen2022universal}. This dataset spans 89 elements and comprises 187,687 snapshots from 62,783 compounds captured during their structural relaxation processes. By excluding compounds missing either initial or DFT-relaxed structures, we refined the dataset to 62,724 pairs. Each pair consists of an initial and a corresponding DFT-relaxed structure, providing a comprehensive basis for assessing the performance of DeepRelax. This dataset is then split into training, validation, and test data in the ratio of 90\%, 5\%, and 5\%, respectively. As illustrated in Suppl. Fig. 1, the structural differences for each pair tend toward an MAE of zero, indicating that many initial structures are closely aligned with their DFT-relaxed counterparts. 

Training a direct ML model for datasets with varied compositions poses significant challenges, as evidenced in Cryslator \cite{kim2023structure}. This model shows reduced prediction performance when trained on the diverse MP database. Despite these challenges, DeepRelax demonstrates its robustness and universality. As indicated in Table \ref{tbl:MP}, DeepRelax significantly surpasses the three baseline models in coordinate prediction, highlighting its effectiveness even in diverse and complex datasets. Fig. \ref{fgr:pred_distribution}(b) shows the MAE distribution for predicted structures compared to the DFT-relaxed ones for the MP dataset, which is less significant compared to the results for the X-Mn-O dataset shown in Fig. \ref{fgr:pred_distribution}(a). This is because many initial structures closely resemble their DFT-relaxed structures in the MP database as evidenced by Suppl. Fig. 1. Consequently, the MP dataset presents a more complex learning challenge for structural relaxation models.

\begin{table}[ht]
\caption{Comparison of results between the proposed DeepRelax and other models on the MP dataset. The performances are evaluated by the MAE of coordinates ($\rm \AA$), bond length ($\rm \AA$), lattice ($\rm \AA$), and cell volume ($\rm \AA^3$) between the predicted and DFT-relaxed structures. The improvement is calculated by comparing DeepRelax with the Dummy model}
\label{tbl:MP}
\centering
\tabcolsep=0.4cm
\begin{tabular}{lllll}
\toprule
Model     & Coordinates & Bond length & Lattice & Cell volume \\ \midrule
Dummy     & 0.095           & 0.112           & 0.072    & 27.0           \\
PAINN     & 0.088           & \textbf{0.082}           & 0.043    & 9.3          \\
EGNN      & 0.086           & 0.086           & 0.043    & \textbf{9.3}          \\
DeepRelax & \textbf{0.066}  & 0.094  & \textbf{0.041}    & 9.6 \\
Improvement & 30.53\%       & 16.07\%         & 43.06\%  & 64.44\% 
\\ \bottomrule
\end{tabular}
\\
\begin{flushleft}
\footnotesize{The best performance in each metric is highlighted in bold.}
\end{flushleft}
\end{table}

\subsection{Transfer learning of pre-trained DeepRelax on Computational 2D Materials Database}
Given that most materials databases do not provide the energy and force information of unrelaxed structures, it is difficult for conventional iterative ML models to transfer the trained model from Materials Project to other materials databases. This difficulty arises because transfer learning typically depends on the availability of energy and force information to fine-tune the model. DeepRelax, with its direct structural prediction feature, is more compatible with transfer learning, making it a flexible tool even when only structural data are available.

To demonstrate the reliable application of DeepRelax, we extend the application of DeepRelax, initially pre-trained on 3D materials from the MP dataset, to 2D materials through transfer learning. We take the C2DB dataset \cite{haastrup2018computational, gjerding2021recent, lyngby2022data} as an example, which covers 62 elements and comprises 11,581 pairs of 2D crystal structures, each consisting of an initial and a DFT-relaxed structure. The dataset is divided into training, validation, and testing subsets, maintaining a ratio of 6:2:2. The structural differences for each pair in this dataset fall within the range observed for the X-Mn-O and MP datasets, as shown in Suppl. Fig. 1.

In this application, DeepRelax trained via transfer learning is denoted as DeepRelaxT to differentiate it from DeepRelax. Table \ref{tbl:C2DB} illustrates our key findings: Firstly, both DeepRelax and DeepRelaxT outperform the other three baselines in the C2DB dataset, proving the applicability of our direct ML model to 2D materials. Fig. \ref{fgr:pred_distribution}(c) presents the MAE distribution for predicted structures by the Dummy model and DeepRelax on the C2DB dataset. These results suggest a modest improvement over the Dummy model. Notably, this improvement surpasses those observed for the MP dataset as depicted in Fig. \ref{fgr:pred_distribution}(b). Secondly, DeepRelaxT demonstrates notable improvements over DeepRelax, with enhancements of 5.61\% in coordinates, 38.43\% in bond length, 3.53\% in lattice, and 5.81\% in cell volume in terms of MAE. Finally, DeepRelaxT shows a faster convergence rate than DeepRelax, as detailed in Suppl. Fig. 2. These results underline the benefits of large-scale pretraining and the efficacy of transfer learning.

\begin{table}[ht]
\caption{Comparison of results among DeepRelax, DeepRelaxT (transfer learning version), and other models on the C2DB dataset. The performances are evaluated by the MAE of coordinates ($\rm \AA$), bond length ($\rm \AA$), lattice ($\rm \AA$), and cell volume ($\rm \AA^3$) between the predicted and DFT-relaxed structures}
\label{tbl:C2DB}
\centering
\tabcolsep=0.35cm
\begin{tabular}{lllll}
\toprule
Model        & Coordinates & Bond length & Lattice  & Cell volume \\ \midrule
Dummy        & 0.268           & 0.400           & 0.142    & 149.6           \\
PAINN        & 0.226           & 0.283           & 0.086    & 61.9          \\
EGNN         & 0.232           & 0.311           & 0.089    & 67.9          \\
DeepRelax    & 0.196           & 0.268           & 0.085    & 60.2            \\
DeepRelaxT   & \textbf{0.185}  & \textbf{0.165}  & \textbf{0.082}    & \textbf{56.7} \\
\bottomrule
\end{tabular}
\\
\begin{flushleft}
\footnotesize{The best performance in each metric is highlighted in bold.}
\end{flushleft}
\end{table}

\subsection{Application in layered vdW crystals}
Layered vdW crystals are of significant interest in the field of materials science and nanotechnology because of their unique tunable structures, such as twisting and sliding configurations \cite{li2022deep}. One notable characteristic of these crystals is that the weak inter-layer vdW force may significantly change upon full relaxation, while the strong intra-layer chemical bonds undergo relatively small changes.

To demonstrate the reliable performance of our DeepRelax model on this type of crystal, we performed DFT relaxation of 58 layered vdW crystals covering 29 elements using van der Waals corrections, parameterized within the DFT-D3 Grimme method. Given the small sample size, we employ transfer learning, utilizing a model pre-trained on the Materials Project dataset.

Suppl. Table 1 shows the inter-layer distances for the unrelaxed, DFT-D3-relaxed, and DeepRelax-predicted structures of six vdW layered crystals. The inter-layer distances of the predicted structures closely match those of the relaxed structures, highlighting the effectiveness of transferred DeepRelax on layered vdW crystals. Furthermore, an analysis of the MAE in bond length for representative bonding pairs, detailed in Suppl. Table 2, further demonstrates DeepRelax's precision in predicting structural changes in layered vdW crystals.

\subsection{Application in crystals with defects}
Most crystals have intrinsic defects. To demonstrate the robustness of DeepRelax to crystal structures with neutral point defects, we employ $\rm MoS_2$ structures with a low defect concentration, including 5,933 different defect configurations within an 8 $\times$ 8 supercell, as cataloged by Huang et al. \cite{huang2023unveiling}, to evaluate DeepRelax. Suppl. Fig. 3 demonstrates a notably lower MAE in both atom coordinates and bond lengths for DeepRelax compared to the Dummy model, thereby underscoring DeepRelax's robustness and efficacy in defect structure calculations, which is further validated by DFT calculations in the next chapter.

\begin{figure}[!h]
  \centering
  \includegraphics[width=11.0cm]{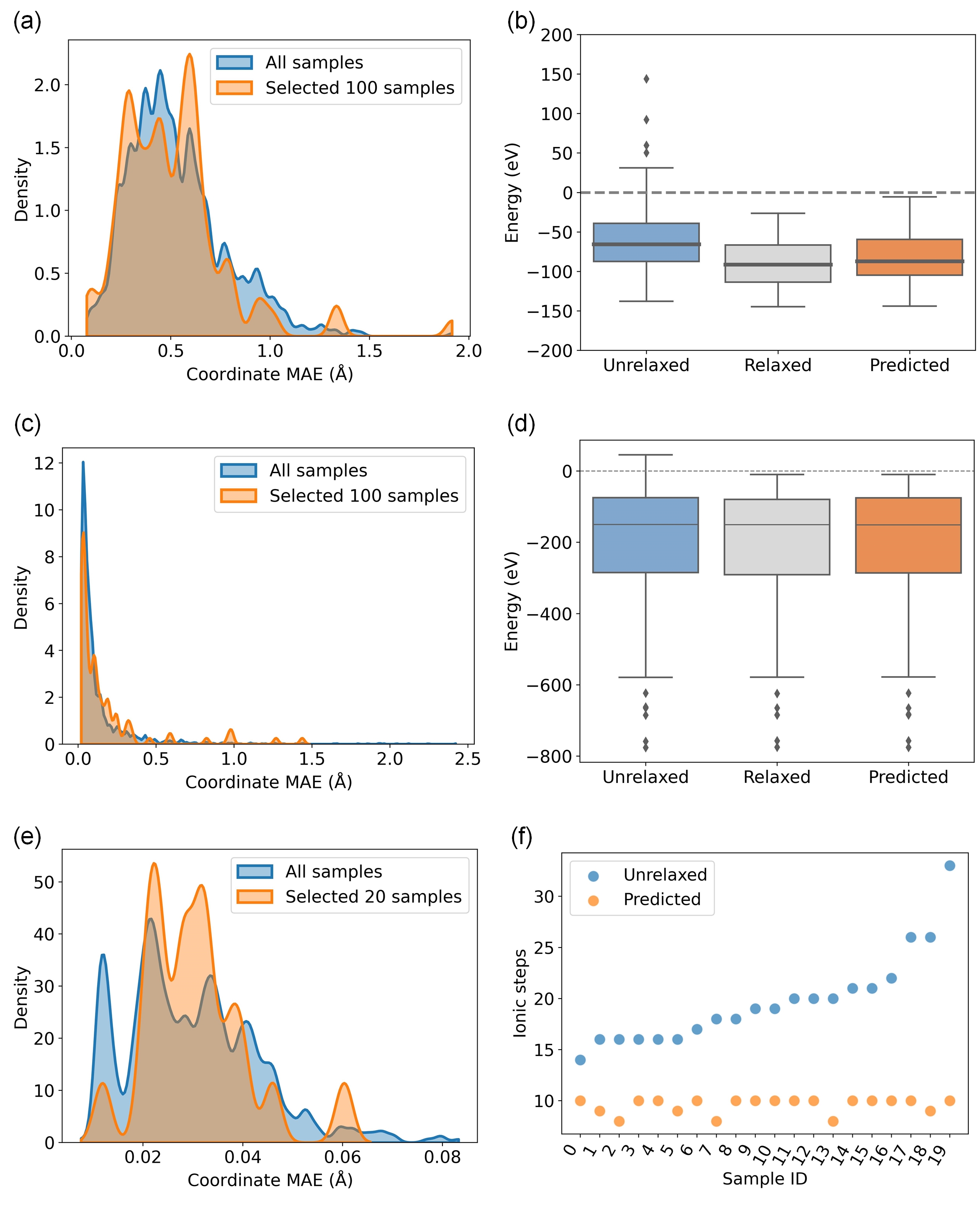}
  \caption{DFT validations. (a) Distributions of deviations for 100 random samples from the X-Mn-O dataset, measured using MAE in coordinates ($\rm \AA$) between the unrelaxed and DFT-relaxed structures. (b) Energy distribution for the three types of structures among the 100 random samples from the X-Mn-O dataset. The boxplots show the median (black line inside the box), interquartile range (box), and whiskers extending to 1.5 times the interquartile range, with outliers plotted as individual points. (c) Distribution of deviations for 100 random samples from the Materials Project dataset with relatively rational initial structures. (d) Energy distribution for the three types of structures among the 100 random samples from the Materials Project dataset. (e) Distributions of deviations for 20 random samples from the 2D materials defect dataset. (f) The number of DFT ionic steps required, starting from the initial unrelaxed structures and the DeepRelax-predicted structures, respectively. The samples are sorted based on the number of ionic steps required by the unrelaxed structures for better observation. Source data are provided with this paper.}
  \label{fgr:DFT}
\end{figure}

\subsection{DFT validations}
\label{sec:DFT_validations}
Usually, the initial crystal structure may deviate from or be close to the final relaxed structure. To demonstrate the efficacy and robustness of DeepRelax, we perform DFT validations on two types of initial structures: those from the X-Mn-O dataset, which exhibit large deviations from the DFT-relaxed state, and those from the MP dataset, which are generally closer to their DFT-relaxed structures, as illustrated in Suppl. Fig. 1. The detailed settings for the DFT calculations are provided in Subsection \ref{sec:dft}.

In the first experiment, we evaluated our model's predictive capability under challenging conditions using the X-Mn-O dataset. We filtered out unrelaxed structures from the X-Mn-O test set that are structurally similar to their DFT-relaxed counterparts using Pymatgen's ``Structure\_matcher" function. From the remaining test set (\(N=1007\)), we randomly selected 100 samples. Fig. \ref{fgr:DFT}(a) shows the deviation distribution for the selected unrelaxed structures, which closely aligns with that of the complete test set, thus confirming the representativeness of the selected subset. Subsequently, we employed DeepRelax to predict the relaxed structures for these samples. Fig. \ref{fgr:DFT}(b) shows box plots of the energy distributions for the unrelaxed, DFT-relaxed, and DeepRelax-predicted structures. The energy distributions of the DeepRelax-predicted and DFT-relaxed structures show similar medians and interquartile ranges, validating the model's accuracy in predicting energetically favorable structures. The MAE in energy is significantly reduced by an order of magnitude from 32.51 to 5.97.

In the second experiment, we tested whether DeepRelax remains effective with structures starting from a relatively rational initial unrelaxed state using the Materials Project dataset. Here, we again randomly selected 100 samples from the test set. Fig. \ref{fgr:DFT}(c) shows the deviation distribution for these samples. The energies of the unrelaxed, DFT-relaxed, and DeepRelax-predicted structures were calculated using DFT. Fig. \ref{fgr:DFT}(d) shows that the predicted structures feature an energy distribution nearly identical to that of the DFT-relaxed structures, demonstrating the model's effectiveness in handling relatively rational initial unrelaxed structures.

Besides the energy indicator, we further demonstrated our model's effectiveness using the number of residual optimizing steps required for DFT relaxation. Specifically, we randomly selected 20 structures from the test set of the point-defect dataset, with their deviation distribution as shown in Fig. \ref{fgr:DFT}(e). We then conducted DFT calculations starting from the unrelaxed and DeepRelax-predicted structures, respectively. As shown in Fig. \ref{fgr:DFT}(f), starting DFT relaxation from the DeepRelax-predicted structures significantly reduces the number of required ionic steps, which is also robust.

\subsection{Analysis of uncertainty}
A critical challenge in integrating artificial intelligence (AI) into material discovery is establishing trustworthy AI models. Current deep learning models typically offer accurate predictions only within the chemical space covered by their training datasets, known as the applicability domain \cite{yu2022uncertainty}. Predictions for samples outside this domain can be questionable. Thus, uncertainty quantification has become critical for AI models by quantifying prediction confidence levels, thereby aiding researchers in decision-making and experimental planning.

\begin{figure}[ht]
  \centering
  \includegraphics[width=11.8cm]{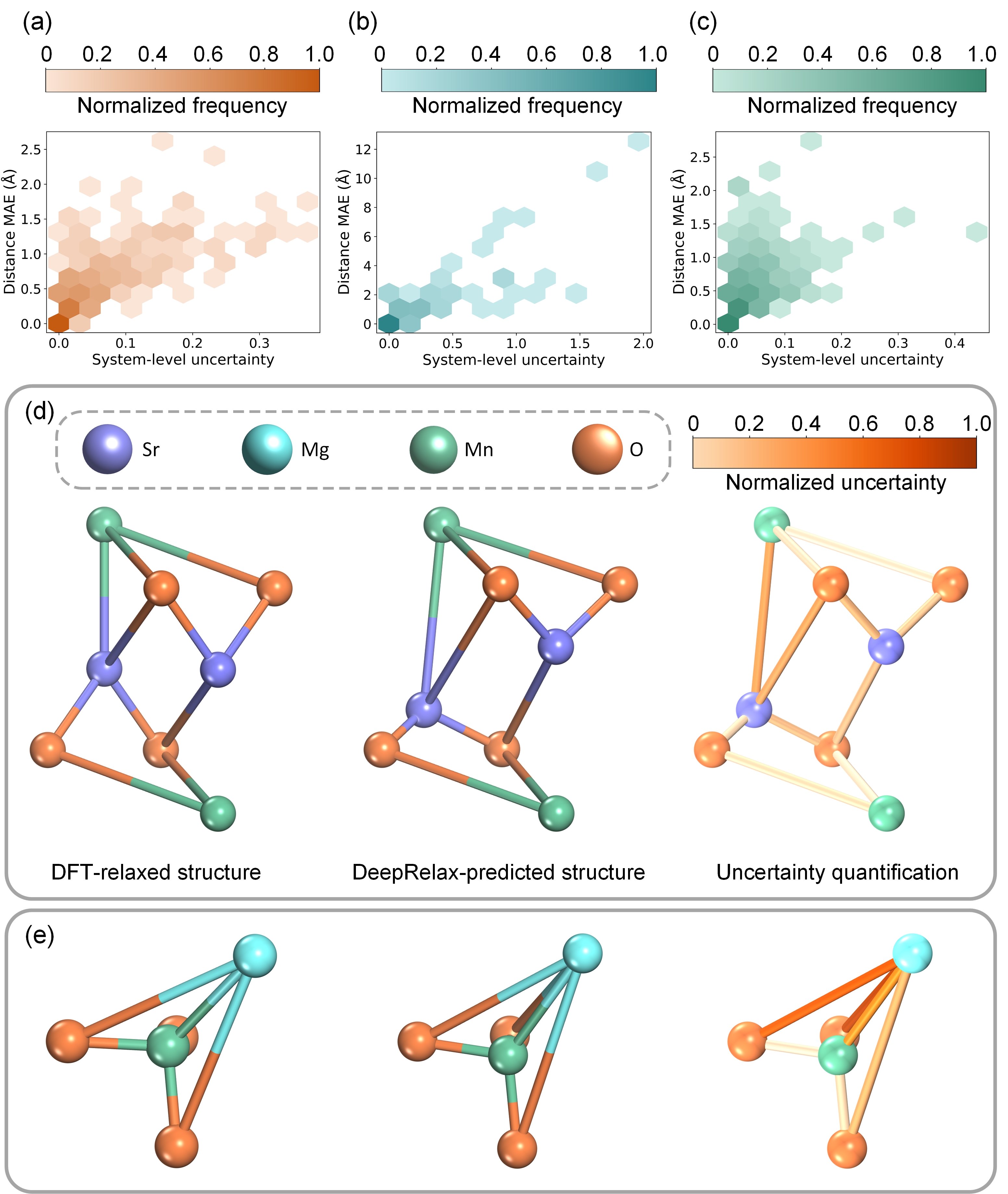}
  \caption{Uncertainty quantification. Hexagonal binning plots comparing system-level uncertainty with distance MAE ($\rm \AA$) for the (a) X-Mn-O, (b) MP, and (c) C2DB datasets. (d) and (e) illustrate the bond-level uncertainty for each predicted pairwise distance in $\rm Sr_2Mn_2O_4$ and $\rm Mg_1Mn_1O_3$, respectively, demonstrating the correlation between distance prediction errors and their associated bond-level uncertainties. Source data are provided with this paper.}
  \label{fgr:uncertainty}
\end{figure}

To validate the efficacy of our proposed uncertainty quantification in reflecting the confidence level of model predictions, we compute Spearman's rank correlation coefficient between the total predicted distance error and its associated system-level uncertainty. Fig. \ref{fgr:uncertainty}(a)-(c) show the hexagonal binning plots of system-level uncertainty against total distance MAE for the X-Mn-O, MP, and C2DB datasets, respectively. Correlation coefficients of 0.95, 0.83, and 0.88 for these datasets demonstrate a strong correlation between predicted error and predicted system-level uncertainty. Fig. \ref{fgr:uncertainty}(d)-(e) present the bond-level uncertainty visualization for two predicted structures, illustrating the correlation between predicted bond length error and associated bond-level uncertainty. These results indicate that the model's predicted uncertainty is a good indicator of the predicted structure's accuracy.

\begin{table}[ht]
\centering
\caption{Ablation study to investigate the impact of unit cell offset encoding (UCOE) and bond-level data uncertainty (BLDU) estimation on model performance. The performances are evaluated by the MAE of coordinates ($\rm \AA$), bond length ($\rm \AA$), lattice ($\rm \AA$), and cell volume ($\rm \AA^3$) between the predicted and DFT-relaxed structures}
\label{tbl:ablation}
\tabcolsep=0.35cm
\begin{tabular}{lllll}
\hline
Model      & Coordinates & Bond length & Lattice & Cell volume \\ \hline
Dummy      & 0.314       & 0.429       & 0.221   & 32.839      \\
Vanilla    & 0.155       & 0.170       & 0.063   & 3.478       \\
DeepRelax (UCOE) & 0.121       & 0.147       & 0.063   & 3.563       \\
DeepRelax (BLDU) & 0.142       & 0.171       & 0.064   & 3.539       \\
DeepRelax  & \textbf{0.116}       & \textbf{0.136}       & \textbf{0.063}   & \textbf{3.442}       \\ \hline
\end{tabular}
\\
\begin{flushleft}
\footnotesize{The best performance in each metric is highlighted in bold.}
\end{flushleft}
\end{table}

\subsection{Ablation study}
DeepRelax's technical contributions are twofold: it utilizes UCOE for handling PBCs explicitly, and it employs a method for estimating bond-level data uncertainty to encourage the model to capture a more comprehensive representation of the underlying data distribution.

To validate the effectiveness of these two strategies, we introduce three additional baseline models for comparison:
\begin{itemize}
    \item Vanilla: Excludes both UCOE and data uncertainty estimation.
    \item DeepRelax (UCOE): Integrates UCOE but omits data uncertainty estimation.
    \item DeepRelax (BLDU): Implements bond-level data uncertainty estimation but not UCOE.
\end{itemize}
Table \ref{tbl:ablation} demonstrates that DeepRelax (UCOE) attains a significant performance enhancement over the Vanilla model, suggesting the UCOE contributes greatly to model performance. On the other hand, DeepRelax (BLDU) shows a more modest improvement, which indicates the added value of data uncertainty estimation. Overall, DeepRelax shows a 25.16\% improvement in coordinate MAE and a 20.00\% advancement in bond length MAE over the Vanilla model. These comparative results underscore the combined effectiveness of UCOE and data uncertainty estimation in our final DeepRelax model.

\section{Discussion}
The rapid advancement of generative models like CDVAE \cite{xie2021crystal}, PGCGM \cite{zhao2023physics}, and MatterGen \cite{zeni2023mattergen}, has opened avenues for the prolific generation of hypothetical materials with potentially desirable properties, such as 2.2 million new materials recently discovered by Google DeepMind. Clearly, it is impossible to relax such a huge number of structures using the traditional ab initio method, and it is also very difficult using the iterative ML relax models. For example, we further compare the efficiency between DeepRelax and M3GNet, a leading iterative ML model. DeepRelax offers a substantial speed advantage, being approximately 100 times faster than M3GNet (see Suppl. Table 3). Based on this estimation, to relax the 2.2 million new materials, our DeepRelax model only needs around 100 hours or 4 days, while M3GNet will take around 400 days. Moreover, our DeepRelax model supports parallel GPU processing, which can further significantly reduce computer time. While there are other direct structure-prediction ML methods, such as DOGSS \cite{yoon2020differentiable} and Cryslator \cite{kim2023structure}, detailed comparisons with these methods are provided in Suppl. Note 6. Overall, we introduce a fast, scalable, and trustworthy deep generative model, DeepRelax, for direct structural relaxation. Despite its advancements, opportunities for further improvement remain, which we explore in subsequent discussions.

Firstly, DeepRelax primarily focuses on predicting interatomic distances, which are quantities fundamentally involving two-body interactions. Incorporating the prediction of higher-order many-body quantities could further enhance the accuracy of structural predictions.

Secondly, implementing active learning strategies \cite{tran2018active, szymanski2023autonomous} may further enhance DeepRelax's performance, particularly in underexplored chemical spaces. Active learning efficiently reduces the need for extensive training data by strategically choosing the most informative samples. DeepRelax's capability to assess prediction uncertainty aligns well with the principles of active learning, suggesting its feasibility as a future enhancement method.

Thirdly, DeepRelax is not designed to replace DFT, but to significantly speed up the traditional ab initio relaxation process, especially for complex structures, such as complex chemical reaction surfaces or doped semiconductor interfaces.

In conclusion, DeepRelax represents a significant advancement in crystal structure prediction, offering efficient, scalable, universal, and trusted structural relaxation capabilities. It excels at direct predictions from initial configurations and effectively handles periodic boundary conditions, along with incorporating uncertainty quantification. DeepRelax thus stands as a powerful tool in advancing material science research.

\section{Methods}
\label{sec:methods}

\subsection{Periodicity in crystals}
A crystal can be conceptualized as a periodic arrangement of atoms in 3D space \cite{xie2021crystal}. This periodicity is typically captured by a unit cell, that effectively represents the crystal structure. Such a unit cell, containing $N$ atoms, can be fully characterized by three components:
\begin{itemize}
    \item Atom Types: Represented by $\bm{A} = (a_0, ..., a_N) \in \mathbb{A}^N$, where $\mathbb{A}$ denotes the set of all chemical elements.
    \item Atom Coordinates: Denoted by $\bm{R} = (\vec{\bm{r}}_0, ..., \vec{\bm{r}}_N) \in \mathbb{R}^{N \times 3}$.
    \item Lattice Vectors: Expressed as $\bm{L} = (\vec{\bm{l}}_1, \vec{\bm{l}}_2, \vec{\bm{l}}_3) \in \mathbb{R}^{3 \times 3}$.
\end{itemize}
Given $\bm{M} = (\bm{A}, \bm{R}, \bm{L})$, we can model the infinite periodic structure as:
\begin{equation}
    \{(a^\prime_i, \vec{\bm{r}^\prime}_i)\vert a^\prime_i = a_i, \vec{\bm{r}^\prime_i} = \vec{\bm{r}}_i + k_1\vec{\bm{l}}_1 + k_2\vec{\bm{l}}_2 + k_3\vec{\bm{l}}_3, k_1, k_2, k_3 \in \mathbb{Z}\},
\end{equation}
where $(k_1, k_2, k_3)$ are the unit cell offsets used to replicate the unit cell across the 3D space.

\begin{figure}[p]
  \centering
  \includegraphics[width=11.8cm]{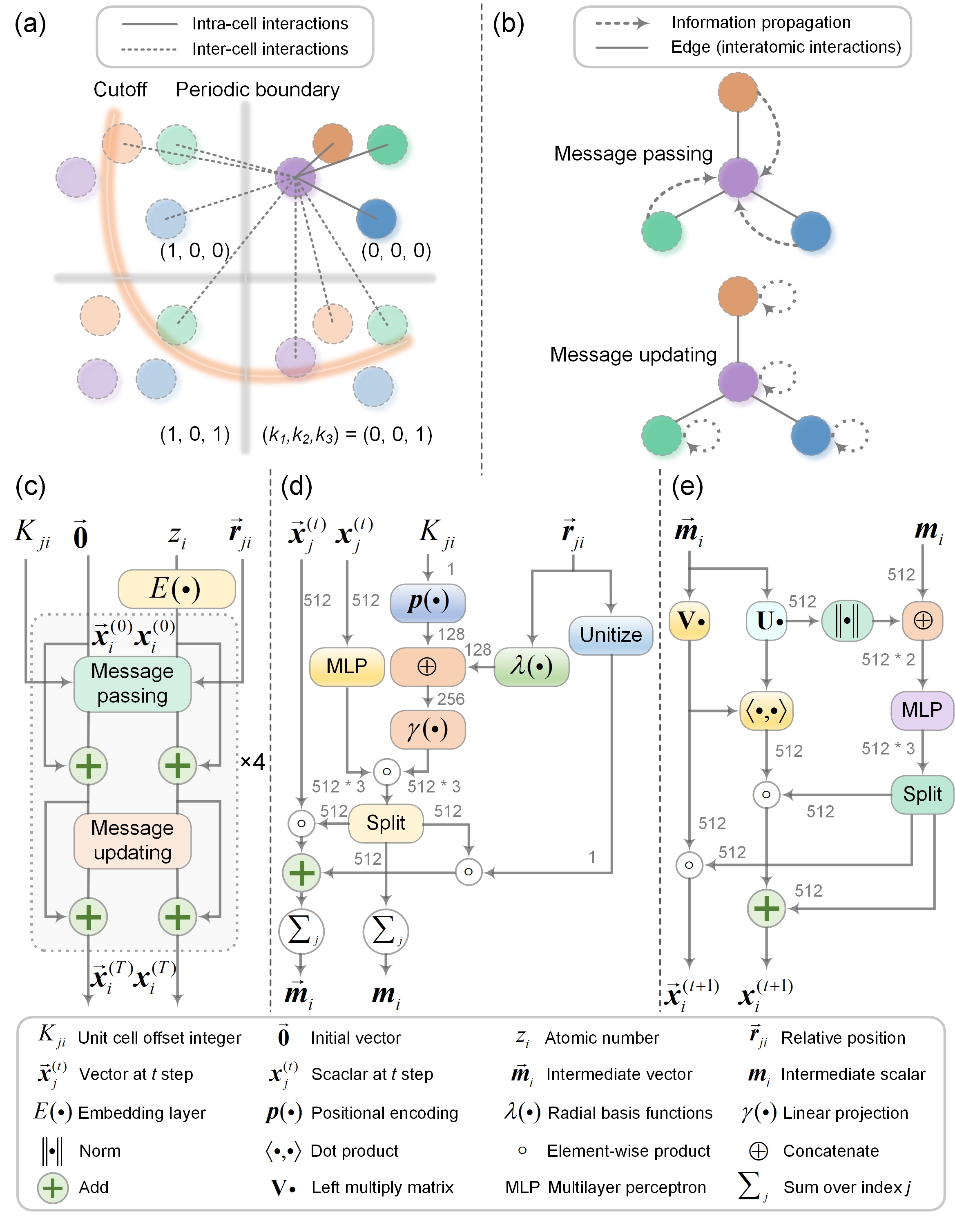}
  \caption{The architecture of PaEGNN. (a) Illustration of the multi-graph representation designed to capture atomic interactions across cell boundaries in periodic structures. (b) Message passing that collects messages from a node's neighbors and message updating that updates node representations using a node's internal states. (c) Overview of PaEGNN, comprising four layers, each with message passing and message updating phases, taking unit cell offset integer $K_{ji}$, initial vector $\vec{\bm{x}}_i^{(0)} = \vec{\bm{0}}$, initial scalar $\bm{x}_i^{(0)} = E(z_i)$, and relative position $\vec{\bm{r}}_{ij}$ as inputs and outputting the final vector $\vec{\bm{x}}_i^{(T)}$ and scalar $\bm{x}_i^{(T)}$ representations. (d) During the message passing phase, a node $v_i$ aggregates messages from neighboring vectors $\vec{\bm{x}}_j^{(t)}$ and scalars $\bm{x}_j^{(t)}$, forming intermediate vector and scalar variables $\vec{\bm{m}}_i$ and $\bm{m}_i$. (e) The message updating phase integrates the $F$ vectors and $F$ scalars within $\vec{\bm{m}}_i$ and $\bm{m}_i$ to generate updated vector $\vec{\bm{x}}_i^{(t+1)}$ and scalar $\bm{x}_i^{(t+1)}$.
  }
  \label{fgr:PaEGNN}
\end{figure}

\subsection{Multi-graph representation for crystal structures}
Multi-graphs offer an intuitive way to represent crystal structures under periodic boundary conditions (PBCs) \cite{xie2021crystal}, as depicted in Fig. \ref{fgr:PaEGNN}(a). These graphs can be effectively processed by GNNs through graph convolutions or message passing, which simulate many-body interactions \cite{li2024jmi, musaelian2023learning, pablo2023fast, li2022deep, gong2023general, zhong2023transferable, zhong2024universal, xie2018crystal, park2020developing, schutt2017schnet, chen2019graph, gasteiger_dimenet_2020, choudhary2021atomistic, chen2022universal, deng2023chgnet, unke2021spookynet,  batzner20223, banik2023cegann, unke2019physnet, zhang2022interpretable,  satorras2021n, omee2022scalable, schutt2021equivariant, haghighatlari2022newtonnet, han2024survey}. Formally, we define a multi-graph $\mathcal{G} = (\mathcal{V}, \mathcal{E})$ to encode these periodic structures. Here, $\mathcal{V}=\{v_1, ..., v_N\}$ represents the set of nodes (atoms), and $\mathcal{E}=\{e_{ij,(k_1, k_2, k_3)}\vert i, j \in \{1, ..., N\}, k_1, k_2, k_3 \in \mathbb{Z}\}$ signifies the set of edges (bonds). The edge $e_{ij,(k_1,k_2,k_3)}$ denotes a directed connection from node $v_i$ in the original unit cell to node $v_j$ in the unit cell translated by $k_1\vec{\bm{l}}_1 + k_2\vec{\bm{l}}_2 + k_3\vec{\bm{l}}_3$. Nodes are interconnected with their nearest neighbors within a cutoff distance $D$ (set to $6 \ \rm \AA$ in our study).

To actively explore the crystal symmetry, each node $v_i \in \mathcal{V}$ is assigned both a scalar feature $\bm{x}_i \in \mathbb{R}^F$ and a vector feature $\vec{\bm{x}}_i \in \mathbb{R}^{F \times 3}$, i.e., retaining $F$ scalars and $F$ vectors for each node. These features are updated in a way that preserves symmetry during training. The scalar feature $\bm{x}_i^{(0)}$ is initialized as an embedding dependent on the atomic number, $E(z_i) \in \mathbb{R}^F$, where $z_i$ is the atomic number, and $E$ is an embedding layer mapping $z_i$ to a $F$-dimensional feature vector. This embedding is similar to the one-hot vector but is trainable. The vector feature is initially set to zero, $\vec{\bm{x}}_i^{(0)}=\vec{\bm{0}} \in \mathbb{R}^{F \times 3}$. Additionally, we define $\vec{\bm{r}}_{ij} = \vec{\bm{r}}_j - \vec{\bm{r}}_i$ as the vector from node $v_i$ to $v_j$.

\subsection{Periodicity-aware equivariant graph neural network}
PaEGNN iteratively updates node representations in two phases: message passing and updating. These phases are illustrated in Fig. \ref{fgr:PaEGNN}(b) and further detailed in Fig. \ref{fgr:PaEGNN}(c)-(e). During message passing, nodes receive information from neighboring nodes, expanding their accessible radius. In the updating phase, PaEGNN utilizes the node's internal messages (composed of $F$ scalars and $F$ vectors) to update its features. To prevent over-smoothing \cite{yang2022learning, yang2022mgraphdta}, skip connections are added to each layer. 

In subsequent sections, we define the norm $\Vert \cdot \Vert$ and dot product $\langle \cdot, \cdot \rangle$ as operations along the spatial dimension, while concatenation $\oplus$ and the element-wise product $\circ$ are performed along the feature dimension.

\subsubsection{Unit cell offset encoding}
A notable feature of PaEGNN, distinguishing it from previous models \cite{schutt2021equivariant, xie2018crystal}, is the explicit differentiation of atoms in various translated unit cells to encode PBCs. To achieve this, we define the set $\mathcal{C}=\{-2, -1, 0, 1, 2\}$. We then use this set to generate translated unit cells with offsets $(k_1, k_2, k_3) \in \mathcal{C} \times \mathcal{C} \times \mathcal{C}$. The translated unit cells, resulting from the offsets $(k_1, k_2, k_3) \in \mathcal{C} \times \mathcal{C} \times \mathcal{C}$, are generally sufficient to encompass all atoms within a $6 \ \text{\AA}$ cutoff distance. We use $(k_1,k_2,k_3)_{ij}$ to denote the unit cell offset from node $v_i$ to node $v_j$, where node $v_j$ is located in a unit cell translated by $k_1\vec{\bm{l}}_1 + k_2\vec{\bm{l}}_2 + k_3\vec{\bm{l}}_3$ relative to node $v_i$. Let $K_{ij}=(k_1+2)+(k_2+2)5+(k_3+2)25$ be a positive integer that uniquely indexes the unit cell offset, the sinusoidal positional encoding \cite{vaswani2017attention} for $K_{ij}$ is computed as:
\begin{equation}
    p(K_{ij}, f) = 
    \begin{cases}
      \sin(K_{ij}/10000^{f/F}),  & \text{if}\ f \in \{0,2,4,\dots,F-2\} \\
      \cos(K_{ij}/10000^{(f-1)/F}),  & \text{if}\ f \in \{1,3,5,\dots,F-1\}
    \end{cases}
\end{equation}
The full positional encoding vector is then 
\begin{equation}
\bm{p}(K_{ij})=\big( p(K_{ij},0), p(K_{ij},1), \dots, p(K_{ij},F-1) \big) \in \mathbb{R}^F
\end{equation}
The unit cell offset encoding $\bm{p}(K_{ij})$ explicitly encodes the relative position of the unit cells in which the two nodes, $v_i$ and $v_j$, are located. This encoding enables the GNN to explicitly recognize the periodic structure, thereby enhancing predictive performance.

\subsubsection{Message passing phase}
During this phase, a node $v_i$ aggregates messages from neighboring scalars $\bm{x}_j^{(t)}$ and vectors $\vec{\bm{x}}_j^{(t)}$, forming intermediate scalar and vector variables $\bm{m}_i$ and $\vec{\bm{m}}_i$ as follows:
\begin{equation}
    \bm{m}_i= \sum_{v_j \in \mathcal{N}(v_i)}( \bm{W}_h \bm{x}_{j}^{(t)}) \circ \gamma_h \big(\lambda(\Vert \vec{\bm{r}}_{ji} \Vert) \oplus \bm{p}(K_{ji})\big)
\end{equation}
\begin{align}
    \vec{\bm{m}}_i = \sum_{v_j \in \mathcal{N}(v_i)} & (\bm{W}_u \bm{x}_{j}^{(t)}) \circ \gamma_u \big(\lambda(\Vert \vec{\bm{r}}_{ji} \Vert) \oplus \bm{p}(K_{ji})\big) \circ \vec{\bm{x}}_j^{(t)} \nonumber \\
    & + (\bm{W}_v \bm{x}_{j}^{(t)}) \circ \gamma_v \big(\lambda(\Vert \vec{\bm{r}}_{ji} \Vert) \oplus \bm{p}(K_{ji})\big) \circ \frac{\vec{\bm{r}}_{ji}}{\Vert \vec{\bm{r}}_{ji} \Vert}
\end{align}
Here, $\oplus$ denotes concatenation, $\mathcal{N}(v_i)$ represents the neighboring nodes of $v_i$, $\bm{W}_h, \bm{W}_u, \bm{W}_v \in \mathbb{R}^{F \times F}$ are trainable weight matrices, $\lambda$ is a set of Gaussian radial basis functions (RBF) \cite{schutt2017schnet} that are used to expand bond distances, and $\gamma_h$, $\gamma_u$, and $\gamma_v$ are a linear projection mapping the concatenated feature back to $F$-dimensional space.

\subsubsection{Message updating phase}
The updating phase concentrates on integrating the $F$ scalars and $F$ vectors within $\bm{m}_i$ and $\vec{\bm{m}}_i$ to generate updated scalar $\bm{x}_i^{(t+1)}$ and vector $\vec{\bm{x}}_i^{(t+1)}$:
\begin{equation}
    \bm{x}_i^{(t+1)}= \bm{W}_{s1}(\bm{m}_{i} \oplus \Vert \bm{U}  \vec{\bm{m}}_i \Vert) + \bm{W}_{s2}(\bm{m}_{i} \oplus \Vert \bm{U}  \vec{\bm{m}}_i \Vert) \langle \bm{V} \vec{\bm{m}}_i, \bm{U}\vec{\bm{m}}_i \rangle
\end{equation}
\begin{equation}
    \vec{\bm{x}}_i^{(t+1)} = \bm{W}_{v}(\bm{m}_{i} \oplus \Vert \bm{U}  \vec{\bm{m}}_i \Vert) \circ (\bm{V} \vec{\bm{m}}_i)
\end{equation}
where $\bm{W}_{s1}, \bm{W}_{s2}, \bm{W}_{v}  \in \mathbb{R}^{F \times 2F}$ and $\bm{U}, \bm{V} \in \mathbb{R}^{F \times F}$.

\subsection{Predicting relaxation quantities}
\label{sec:relaxation_quantities}
Assuming PaEGNN comprises $T$ layers, and we define bond feature $\bm{h}_{ij}=\gamma \big(\lambda(\Vert \vec{\bm{r}}_{ij} \Vert) \oplus \bm{p}(k_1,k_2,k_3)\big)$, where $\gamma$ is a linear projection mapping the concatenated feature back to $F$-dimensional space. The prediction of a pairwise distance $\hat{d}_{ij}$ for the edge $e_{ij, (k_1,k_2,k_3)}$ is formulated as:
\begin{equation} \label{eqn:distance}
    \hat{d}_{ij} = \vert f_d(\bm{W}_d\bm{x}_i^{(T)} \oplus \bm{W}_d\bm{x}_j^{(T)} \oplus \bm{h}_{ij}) \vert 
\end{equation}
where $\bm{W}_d \in \mathbb{R}^{F \times F}$ is a learnable matrix, and $f_d: \mathbb{R}^{3F} \rightarrow \mathbb{R}$ is linear maps. Using Eqn. \eqref{eqn:distance}, we can predict both the interatomic distances in the relaxed structure and the displacements between the initial and relaxed structures. Additionally, DeepRelax predicts the lattice matrix of the relaxed structure as follows:
\begin{equation}
\hat{\bm{L}}=r_L\Big( f_L\big (\bm{W}_l(\vec{\bm{l}}_1 \oplus \vec{\bm{l}}_2 \oplus \vec{\bm{l}}_3) \oplus (\sum_{v_i \in \mathcal{G}} \bm{W}_x \bm{x}_i) \big) \Big)
\end{equation}
Here, $\bm{W}_x \in \mathbb{R}^{F \times F}$, $\bm{W}_l \in \mathbb{R}^{9 \times F}$, and $f_L: \mathbb{R}^{2F}\rightarrow \mathbb{R}^9$ is a linear mapping yielding a 9-dimensional vector $\bm{L}_v$. The operation $r_L$ reshapes $\bm{L}_v$ into a $3 \times 3$ matrix $\hat{\bm{L}}$ to reflect the lattice vectors.

\subsection{Uncertainty-aware loss function}
In real scenarios, each predicted distance is subject to inherent noise (e.g., measurement errors or human labeling errors). To capture this uncertainty, we can model the pairwise distances as random variables following a Laplace distribution, i.e., $d_{ij} \sim \text{Laplace}(\hat{d}_{ij}, \hat{b}_{ij})$. Here, $\hat{d}_{ij}$ and $\hat{b}_{ij}$ are the location parameter and scale parameter, respectively. In our application, \(\hat{d}_{ij}\) represents the predicted distance, and \(\hat{b}_{ij}\) represents the associated bond-level data uncertainty. The scale parameter \(\hat{b}_{ij}\) is predicted as follows:
\begin{equation}
    \hat{b}_{ij}=f_b(\bm{W}_b\bm{x}_i^{(T)} \oplus \bm{W}_b\bm{x}_j^{(T)} \oplus \bm{h}_{ij})
\end{equation}
where \(\bm{W}_b \in \mathbb{R}^{F \times F}\) is a learnable matrix, and \( f_b: \mathbb{R}^{3F} \rightarrow \mathbb{R} \) is a linear map.

To train DeepRelax such that its output follows the assumed Laplace distribution, we propose an uncertainty-aware loss $\mathcal{L}_u$, which comprises interatomic distance loss \(\mathcal{L}_i\) and displacement loss \(\mathcal{L}_d\):
\begin{equation}
\mathcal{L}_{i}=\sum_{e_{ij,(k_1,k_2,k_3)} \in \mathcal{E}} \log(2\hat{b}_{ij})+\frac{\vert d_{ij} - \hat{d}_{ij} \vert}{\hat{b}_{ij}}
\end{equation}
\begin{equation}
\mathcal{L}_{d}=\sum_{e_{ij,(0,0,0)}\in \mathcal{E}}\log(2\hat{b}_{ij})+\frac{\vert d_{ij} - \hat{d}_{ij} \vert}{\hat{b}_{ij}}
\end{equation}
\begin{equation} \label{eqn:loss}
\mathcal{L}_u=\mathcal{L}_{i}+\mathcal{L}_{d}
\end{equation}
In these expressions, $d_{ij}$ represents the ground truth distance. The edges $e_{ij,(k_1,k_2,k_3)} \in \mathcal{E}$ pertain to interatomic distance predictions, whereas $e_{ij,(0,0,0)}\in \mathcal{E}$ denotes edges used for displacement predictions within the unit cell, discounting PBCs. In essence, \(\mathcal{L}_{i}\) and \(\mathcal{L}_{d}\) represent the negative log-likelihood of the Laplace distribution, thereby capturing the data uncertainty. Consequently, a larger $\hat{b}_{ij}$ indicates greater bond-level data uncertainty in the prediction, and vice versa. The total loss $\mathcal{L}$ is consist of $\mathcal{L}_u$ and a lattice loss $\mathcal{L}_l$:
\begin{equation} 
\mathcal{L}_{l}=\sum \vert \hat{\bm{L}}-\bm{L}\vert
\end{equation}
\begin{equation} \label{eqn:loss}
\mathcal{L}=\mathcal{L}_{u}+\mathcal{L}_{l}
\end{equation}
where $\bm{L}$ represents the ground lattice matrix of the relaxed structure. 

\subsection{Numerical Euclidean distance geometry solver}
We propose a numerical EDG solver to determine the relaxed structure that aligns with the predicted relaxation quantities. Specifically, for a given graph $\mathcal{G}=(\mathcal{V}, \mathcal{E},d)$ and a dimension $K$, the EDG problem \cite{liberti2014euclidean, lu2022tankbind, masters2023deep} seeks a realization—specifically, a coordinate matrix $\bm{\hat{R}}\in \mathbb{R}^{N \times K}:\mathcal{V}\rightarrow \mathbb{R}^K$ in $K$-dimensional space that satisfy the distance constraint $d$ as follows:
\begin{equation}
\forall (u,v)\in \mathcal{E}, \Vert \bm{\hat{R}}(u)-\bm{\hat{R}}(v) \Vert=d_{uv}
\end{equation}
For simplicity in notation, $\bm{\hat{R}}(u)$ and $\bm{\hat{R}}(v)$ are typically written as $\bm{\hat{R}}_u$ and $\bm{\hat{R}}_v$.

We reformulate the conventional EDG problem into a global optimization task:
\begin{equation}
\mathcal{L}_g=\sum_{(u,v)\in \mathcal{E}}\lvert \lVert \bm{\hat{R}}_u-\bm{\hat{R}}_v \rVert-d_{uv} \rvert 
\end{equation}
This is a non-convex optimization problem and minimizing $\mathcal{L}_g$ gives an approximation solution of $\bm{\hat{R}}$.

In our specific application, we aim to find a coordinate matrix $\bm{\hat{R}} \in \mathbb{R}^{N \times 3}$ for a system of $N$ atoms in three-dimensional space, meeting the constraints imposed by $\hat{d}_{ij}$, $\hat{b}_{ij}$, and $\hat{\bm{L}}$. Specifically, we first define an upper bound and a lower bound using $\hat{d}_{ij}$, $\hat{b}_{ij}$ as follows:
\begin{equation} 
    \hat{d}_{ij}^u=\hat{d}_{ij}+\exp(\hat{b}_{ij})
\end{equation}
\begin{equation} 
    \hat{d}_{ij}^l=\hat{d}_{ij}-\exp(\hat{b}_{ij})
\end{equation}
Subsequently, we propose minimizing a bounded Euclidean distance (BED) loss:
\begin{equation}
\mathcal{L}_g = \sum_{\substack{e_{ij, (k_1,k_2,k_3)}\in \mathcal{E} \\ e_{ij,(0,0,0)}\in \mathcal{E}}} \max(0, \lVert \bm{\hat{R}}_u - \bm{\hat{R}}_v \rVert - \hat{d}_{ij}^u) + \sum_{\substack{e_{ij, (k_1,k_2,k_3)}\in \mathcal{E} \\ e_{ij,(0,0,0)}\in \mathcal{E}}} \max(0, \hat{d}_{ij}^l - \lVert \bm{\hat{R}}_u - \bm{\hat{R}}_v \rVert)
\end{equation}
For each edge $e_{ij, (k_1,k_2,k_3)}$, the location of node $v_j$ is dictated by $k_1\hat{\bm{l}}_1 + k_2\hat{\bm{l}}_2 + k_3\hat{\bm{l}}_3$, where $\hat{\bm{l}}_1$, $\hat{\bm{l}}_2$, $\hat{\bm{l}}_3$ are predicted lattice vectors. The BED loss only penalizes coordinate pairs whose distances fall outside the lower and upper bounds, thus mitigating the impact of less accurate predictions. In our work, we use Adam optimizer to minimize $\mathcal{L}_g$.

\subsection{Uncertainty quantification}
We initially quantify bond-level uncertainties and subsequently aggregate these to determine the system-level uncertainty of the predicted structure. The bond-level uncertainty can be further decomposed into data uncertainty and model uncertainty. Data uncertainty arises from the inherent randomness in the data, while model uncertainty arises from a lack of knowledge about the best model to describe the data \cite{gawlikowski2023survey}. 

We employ ensemble-based uncertainty techniques \cite{yu2022uncertainty, luo2023calibrated}, which involve training an ensemble of \( T \) independent model replicates, with \( T=5 \) used in this study. The \( T \) model replicates have the same neural network architectures and hyperparameters, but the learnable parameters are initialized with different random seed. For the \(t\)-th model replicate, let \(\hat{d}_{ij}(t)\) denote the predicted distance, \(\hat{b}_{ij}(t)\) the associated bond-level data uncertainty, and \(\hat{w}_{ij}(t)\) the associated bond-level model uncertainty. Model uncertainty for each pair is calculated as the deviation from the mean predicted distance $\bar{d}_{ij}$:
\begin{equation}
\hat{w}_{ij}(t) = \lvert \hat{d}_{ij}(t)-\bar{d}_{ij}\rvert
\end{equation}
where the mean predicted distance \(\bar{d}_{ij}\) is given by:
\begin{equation}
\bar{d}_{ij} = \frac{1}{T} \sum_{t=1}^T\hat{d}_{ij}(t)
\end{equation}
The total bond-level uncertainty \(\hat{U}_{ij}\) is the sum of the exponential of the data uncertainties and the model uncertainties across $T$ models:
\begin{equation}
    \hat{U}_{ij} = \frac{1}{T} \sum_{t=1}^T \Big( \exp(\hat{b}_{ij}(t)) + \hat{w}_{ij}(t) \Big)
\end{equation}
Finally, the system-level uncertainty \(\hat{U}\) is computed as the average of all bond-level uncertainties:
\begin{equation}
    \hat{U} = \frac{1}{N} \sum_{\substack{e_{ij, (k_1, k_2, k_3)} \in \mathcal{E} \\ e_{ij, (0, 0, 0)} \in \mathcal{E}}} \hat{U}_{ij}
\end{equation}
where \(N\) represents the total number of evaluated pairs.

\subsection{Implementation details}
\label{sec:implementation}
The DeepRelax model is implemented using PyTorch. Experiments are conducted on an NVIDIA RTX A6000 with 48 GB of memory. The training objective is to minimize Eqn. \eqref{eqn:loss}. We use the AdamW optimizer with a learning rate of 0.0001 to update the model's parameters. Additionally, we implement a learning rate decay strategy, reducing the learning rate if there is no improvement in a specified metric for a duration of 5 epochs.

We implement PAINN \cite{schutt2021equivariant} and EGNN \cite{satorras2021n} models, utilizing the source code available at \url{https://github.com/Open-Catalyst-Project/ocp} and \url{https://github.com/vgsatorras/egnn}, respectively. These equivariant models are adept at directly predicting the coordinates of a relaxed structure from its unrelaxed counterpart, leveraging the intrinsic property that coordinates are equivariant quantities.

\subsection{DFT calculations}
\label{sec:dft}
In our study, DFT calculations are performed using the Vienna Ab initio Simulation Package (VASP) \cite{kresse1996efficient}, employing the generalized gradient approximation (GGA) with the Perdew-Burke-Ernzerhof (PBE) exchange-correlation functional. All VASP calculations are performed using the electronic minimization algorithm “all band simultaneous update of orbitals” (ALGO=All), with a cut-off energy of 550 eV, an energy convergence criterion of \(1.0 \times 10^{-5}\) eV, and a Gaussian smearing width of 0.02 eV. For the X-Mn-O dataset, we run the self-consistent calculation to obtain the total energy without spin polarization. The K-point mesh is a $9 \times 9 \times 9$ grid, ensuring precise total energy calculations. The effective on-site Coulomb interactions (U value) of Mn 3d orbital is chosen as 3.9 eV, aligning with that used in Cryslator \cite{kim2023structure}. For the MP dataset, the self-consistent is running with a $5\times 5 \times 5$ K-point mesh for structures containing fewer than 60 atoms and $3\times 3 \times 3$ for those with more than 60 atoms. Spin polarization is applied to structures exhibiting magnetism to enhance the convergence of total energy calculations. For layered vdW crystals, we performed DFT calculations with van der Waals corrections (DFT-D3 Grimme method). For $\rm MoS_2$ structures with defects, the structure is relaxed until the interatomic force is smaller than $0.05 \ \rm{eV /\AA}$. Spin polarization is included following previous studies \cite{huang2023unveiling, kazeev2023sparse}. These high-throughput self-consistent and structural relaxation calculations are implemented utilizing the AiiDA computational framework \cite{uhrin2021workflows}.

\subsection{Performance indicators}
\label{sec:performance_indicators}
\subsubsection{MAE of coordinate}
The MAE of coordinates assesses the structural difference between the predicted and DFT-relaxed structures. It is defined as:
\begin{equation}
\Delta_{\rm coord} = \frac{1}{N}\sum_{v_i \in \mathcal{G}}\lvert \hat{\bm{r}}_i - \vec{\bm{r}}_i \rvert  
\end{equation}
where $N$ represents the total number of nodes in $\mathcal{G}$, $\hat{\bm{r}}_i$ and  $\vec{\bm{r}}_i$ represent the predicted and ground truth Cartesian coordinates, respectively.

\subsubsection{MAE of bond length}
The MAE of bond length measures the error in predicting interatomic distances:
\begin{equation}
\Delta_{\rm bond}=\frac{1}{M} \sum_{e_{ij,(k_1,k_2,k_3)} \in \mathcal{E}} \vert \hat{d}_{ij}-d_{ij} \vert
\end{equation}
where $M$ is the total number of chemical bonds, $\hat{d}_{ij}$ and $d_{ij}$ are the predicted and ground interatomic distances.

\subsubsection{MAE of lattice}
This metric calculates the error in predicting the lattice matrices:
\begin{equation} 
\Delta_{\rm lattice}=\frac{1}{9} \sum \vert \hat{\bm{L}}-\bm{L}\vert
\end{equation}
where $\hat{\bm{L}}$ and $\bm{L}$ are the predicted and ground lattice matrices.

\subsubsection{MAE of cell volume}
The error in predicting the cell volume is given by:
\begin{equation} 
\Delta_{\rm volume}=\Bigl\lvert \vert \hat{\bm{l}}_1 \cdot (\hat{\bm{l}}_2 \times \hat{\bm{l}}_3) \vert - \vert \vec{\bm{l}}_1 \cdot (\vec{\bm{l}}_2 \times \vec{\bm{l}}_3) \vert \Bigr\rvert 
\end{equation}
where $\times$ is the cross product, and $\hat{\bm{l}}_i$ and $\vec{\bm{l}}_i$ are the predicted and ground truth lattice vectors.

\subsubsection{Match rate}
We utilize the ``Structure\_matcher" function from the Pymatgen package \cite{ong2013python} to compare the predicted structure with the DFT-relaxed structure. Default parameters are used for this function (ltol=0.2, stol=0.3) to ensure consistent and objective comparisons.

\section*{Data availability}
The dataset for X-Mn-O is available at \url{https://zenodo.org/records/8081655} (ref. \cite{sungwon_kim_2023_8081655}). The dataset for Materials Project is available at \url{https://figshare.com/articles/dataset/MPF_2021_2_8/19470599} (ref. \cite{Chen2022_19470599}). The dataset for C2DB is available at \url{https://cmr.fysik.dtu.dk/c2db/c2db.html}. The dataset for MoS2 structures with defects is available at \url{https://research.constructor.tech/p/2d-defects-prediction}. The layered vdW crystals dataset is in-house collected and currently unpublished; access can be obtained by contacting Dr. Lei Shen with reasonable requests. Source data and a Python script to reproduce the figures in this paper are provided.

\section*{Code availability}
Code for DeepRelax is available at \url{https://github.com/Shen-Group/DeepRelax} and \url{https://zenodo.org/records/13160937} (ref. \cite{yang_2024_13160937}).


\bibliography{sn-bibliography}


\begin{thebibliography}{70}
\ifx \bisbn   \undefined \def \bisbn  #1{ISBN #1}\fi
\ifx \binits  \undefined \def \binits#1{#1}\fi
\ifx \bauthor  \undefined \def \bauthor#1{#1}\fi
\ifx \batitle  \undefined \def \batitle#1{#1}\fi
\ifx \bjtitle  \undefined \def \bjtitle#1{#1}\fi
\ifx \bvolume  \undefined \def \bvolume#1{\textbf{#1}}\fi
\ifx \byear  \undefined \def \byear#1{#1}\fi
\ifx \bissue  \undefined \def \bissue#1{#1}\fi
\ifx \bfpage  \undefined \def \bfpage#1{#1}\fi
\ifx \blpage  \undefined \def \blpage #1{#1}\fi
\ifx \burl  \undefined \def \burl#1{\textsf{#1}}\fi
\ifx \doiurl  \undefined \def \doiurl#1{\url{https://doi.org/#1}}\fi
\ifx \betal  \undefined \def \betal{\textit{et al.}}\fi
\ifx \binstitute  \undefined \def \binstitute#1{#1}\fi
\ifx \binstitutionaled  \undefined \def \binstitutionaled#1{#1}\fi
\ifx \bctitle  \undefined \def \bctitle#1{#1}\fi
\ifx \beditor  \undefined \def \beditor#1{#1}\fi
\ifx \bpublisher  \undefined \def \bpublisher#1{#1}\fi
\ifx \bbtitle  \undefined \def \bbtitle#1{#1}\fi
\ifx \bedition  \undefined \def \bedition#1{#1}\fi
\ifx \bseriesno  \undefined \def \bseriesno#1{#1}\fi
\ifx \blocation  \undefined \def \blocation#1{#1}\fi
\ifx \bsertitle  \undefined \def \bsertitle#1{#1}\fi
\ifx \bsnm \undefined \def \bsnm#1{#1}\fi
\ifx \bsuffix \undefined \def \bsuffix#1{#1}\fi
\ifx \bparticle \undefined \def \bparticle#1{#1}\fi
\ifx \barticle \undefined \def \barticle#1{#1}\fi
\bibcommenthead
\ifx \bconfdate \undefined \def \bconfdate #1{#1}\fi
\ifx \botherref \undefined \def \botherref #1{#1}\fi
\ifx \url \undefined \def \url#1{\textsf{#1}}\fi
\ifx \bchapter \undefined \def \bchapter#1{#1}\fi
\ifx \bbook \undefined \def \bbook#1{#1}\fi
\ifx \bcomment \undefined \def \bcomment#1{#1}\fi
\ifx \oauthor \undefined \def \oauthor#1{#1}\fi
\ifx \citeauthoryear \undefined \def \citeauthoryear#1{#1}\fi
\ifx \endbibitem  \undefined \def \endbibitem {}\fi
\ifx \bconflocation  \undefined \def \bconflocation#1{#1}\fi
\ifx \arxivurl  \undefined \def \arxivurl#1{\textsf{#1}}\fi
\csname PreBibitemsHook\endcsname

\bibitem{zuo2021accelerating}
\begin{barticle}
\bauthor{\bsnm{Zuo}, \binits{Y.}},
\bauthor{\bsnm{Qin}, \binits{M.}},
\bauthor{\bsnm{Chen}, \binits{C.}},
\bauthor{\bsnm{Ye}, \binits{W.}},
\bauthor{\bsnm{Li}, \binits{X.}},
\bauthor{\bsnm{Luo}, \binits{J.}},
\bauthor{\bsnm{Ong}, \binits{S.P.}}:
\batitle{Accelerating materials discovery with bayesian optimization and graph deep learning}.
\bjtitle{Materials Today}
\bvolume{51},
\bfpage{126}--\blpage{135}
(\byear{2021})
\end{barticle}
\endbibitem

\bibitem{merchant2023scaling}
\begin{barticle}
\bauthor{\bsnm{Merchant}, \binits{A.}},
\bauthor{\bsnm{Batzner}, \binits{S.}},
\bauthor{\bsnm{Schoenholz}, \binits{S.S.}},
\bauthor{\bsnm{Aykol}, \binits{M.}},
\bauthor{\bsnm{Cheon}, \binits{G.}},
\bauthor{\bsnm{Cubuk}, \binits{E.D.}}:
\batitle{Scaling deep learning for materials discovery}.
\bjtitle{Nature}
\bvolume{624}(\bissue{7990}),
\bfpage{80}--\blpage{85}
(\byear{2023})
\end{barticle}
\endbibitem

\bibitem{ong2013python}
\begin{barticle}
\bauthor{\bsnm{Ong}, \binits{S.P.}},
\bauthor{\bsnm{Richards}, \binits{W.D.}},
\bauthor{\bsnm{Jain}, \binits{A.}},
\bauthor{\bsnm{Hautier}, \binits{G.}},
\bauthor{\bsnm{Kocher}, \binits{M.}},
\bauthor{\bsnm{Cholia}, \binits{S.}},
\bauthor{\bsnm{Gunter}, \binits{D.}},
\bauthor{\bsnm{Chevrier}, \binits{V.L.}},
\bauthor{\bsnm{Persson}, \binits{K.A.}},
\bauthor{\bsnm{Ceder}, \binits{G.}}:
\batitle{Python materials genomics (pymatgen): A robust, open-source python library for materials analysis}.
\bjtitle{Computational Materials Science}
\bvolume{68},
\bfpage{314}--\blpage{319}
(\byear{2013})
\end{barticle}
\endbibitem

\bibitem{saal2013materials}
\begin{barticle}
\bauthor{\bsnm{Saal}, \binits{J.E.}},
\bauthor{\bsnm{Kirklin}, \binits{S.}},
\bauthor{\bsnm{Aykol}, \binits{M.}},
\bauthor{\bsnm{Meredig}, \binits{B.}},
\bauthor{\bsnm{Wolverton}, \binits{C.}}:
\batitle{Materials design and discovery with high-throughput density functional theory: the open quantum materials database (oqmd)}.
\bjtitle{Jom}
\bvolume{65},
\bfpage{1501}--\blpage{1509}
(\byear{2013})
\end{barticle}
\endbibitem

\bibitem{curtarolo2012aflow}
\begin{barticle}
\bauthor{\bsnm{Curtarolo}, \binits{S.}},
\bauthor{\bsnm{Setyawan}, \binits{W.}},
\bauthor{\bsnm{Hart}, \binits{G.L.}},
\bauthor{\bsnm{Jahnatek}, \binits{M.}},
\bauthor{\bsnm{Chepulskii}, \binits{R.V.}},
\bauthor{\bsnm{Taylor}, \binits{R.H.}},
\bauthor{\bsnm{Wang}, \binits{S.}},
\bauthor{\bsnm{Xue}, \binits{J.}},
\bauthor{\bsnm{Yang}, \binits{K.}},
\bauthor{\bsnm{Levy}, \binits{O.}}, \betal:
\batitle{Aflow: An automatic framework for high-throughput materials discovery}.
\bjtitle{Computational Materials Science}
\bvolume{58},
\bfpage{218}--\blpage{226}
(\byear{2012})
\end{barticle}
\endbibitem

\bibitem{zhou20192dmatpedia}
\begin{barticle}
\bauthor{\bsnm{Zhou}, \binits{J.}},
\bauthor{\bsnm{Shen}, \binits{L.}},
\bauthor{\bsnm{Costa}, \binits{M.D.}},
\bauthor{\bsnm{Persson}, \binits{K.A.}},
\bauthor{\bsnm{Ong}, \binits{S.P.}},
\bauthor{\bsnm{Huck}, \binits{P.}},
\bauthor{\bsnm{Lu}, \binits{Y.}},
\bauthor{\bsnm{Ma}, \binits{X.}},
\bauthor{\bsnm{Chen}, \binits{Y.}},
\bauthor{\bsnm{Tang}, \binits{H.}}, \betal:
\batitle{2dmatpedia, an open computational database of two-dimensional materials from top-down and bottom-up approaches}.
\bjtitle{Scientific data}
\bvolume{6}(\bissue{1}),
\bfpage{86}
(\byear{2019})
\end{barticle}
\endbibitem

\bibitem{chen2021phase}
\begin{barticle}
\bauthor{\bsnm{Chen}, \binits{B.}},
\bauthor{\bsnm{Conway}, \binits{L.J.}},
\bauthor{\bsnm{Sun}, \binits{W.}},
\bauthor{\bsnm{Kuang}, \binits{X.}},
\bauthor{\bsnm{Lu}, \binits{C.}},
\bauthor{\bsnm{Hermann}, \binits{A.}}:
\batitle{Phase stability and superconductivity of lead hydrides at high pressure}.
\bjtitle{Physical Review B}
\bvolume{103}(\bissue{3}),
\bfpage{035131}
(\byear{2021})
\end{barticle}
\endbibitem

\bibitem{xie2021crystal}
\begin{botherref}
\oauthor{\bsnm{Xie}, \binits{T.}},
\oauthor{\bsnm{Fu}, \binits{X.}},
\oauthor{\bsnm{Ganea}, \binits{O.-E.}},
\oauthor{\bsnm{Barzilay}, \binits{R.}},
\oauthor{\bsnm{Jaakkola}, \binits{T.}}:
Crystal diffusion variational autoencoder for periodic material generation.
arXiv preprint arXiv:2110.06197
(2021)
\end{botherref}
\endbibitem

\bibitem{zeni2023mattergen}
\begin{botherref}
\oauthor{\bsnm{Zeni}, \binits{C.}},
\oauthor{\bsnm{Pinsler}, \binits{R.}},
\oauthor{\bsnm{Z{\"u}gner}, \binits{D.}},
\oauthor{\bsnm{Fowler}, \binits{A.}},
\oauthor{\bsnm{Horton}, \binits{M.}},
\oauthor{\bsnm{Fu}, \binits{X.}},
\oauthor{\bsnm{Shysheya}, \binits{S.}},
\oauthor{\bsnm{Crabb{\'e}}, \binits{J.}},
\oauthor{\bsnm{Sun}, \binits{L.}},
\oauthor{\bsnm{Smith}, \binits{J.}}, et al.:
Mattergen: a generative model for inorganic materials design.
arXiv preprint arXiv:2312.03687
(2023)
\end{botherref}
\endbibitem

\bibitem{zhao2023physics}
\begin{barticle}
\bauthor{\bsnm{Zhao}, \binits{Y.}},
\bauthor{\bsnm{Siriwardane}, \binits{E.M.D.}},
\bauthor{\bsnm{Wu}, \binits{Z.}},
\bauthor{\bsnm{Fu}, \binits{N.}},
\bauthor{\bsnm{Al-Fahdi}, \binits{M.}},
\bauthor{\bsnm{Hu}, \binits{M.}},
\bauthor{\bsnm{Hu}, \binits{J.}}:
\batitle{Physics guided deep learning for generative design of crystal materials with symmetry constraints}.
\bjtitle{npj Computational Materials}
\bvolume{9}(\bissue{1}),
\bfpage{38}
(\byear{2023})
\end{barticle}
\endbibitem

\bibitem{chen2022universal}
\begin{barticle}
\bauthor{\bsnm{Chen}, \binits{C.}},
\bauthor{\bsnm{Ong}, \binits{S.P.}}:
\batitle{A universal graph deep learning interatomic potential for the periodic table}.
\bjtitle{Nature Computational Science}
\bvolume{2}(\bissue{11}),
\bfpage{718}--\blpage{728}
(\byear{2022})
\end{barticle}
\endbibitem

\bibitem{deng2023chgnet}
\begin{botherref}
\oauthor{\bsnm{Deng}, \binits{B.}},
\oauthor{\bsnm{Zhong}, \binits{P.}},
\oauthor{\bsnm{Jun}, \binits{K.}},
\oauthor{\bsnm{Riebesell}, \binits{J.}},
\oauthor{\bsnm{Han}, \binits{K.}},
\oauthor{\bsnm{Bartel}, \binits{C.J.}},
\oauthor{\bsnm{Ceder}, \binits{G.}}:
Chgnet as a pretrained universal neural network potential for charge-informed atomistic modelling.
Nature Machine Intelligence,
1--11
(2023)
\end{botherref}
\endbibitem

\bibitem{mosquera2024machine}
\begin{barticle}
\bauthor{\bsnm{Mosquera-Lois}, \binits{I.}},
\bauthor{\bsnm{Kavanagh}, \binits{S.R.}},
\bauthor{\bsnm{Ganose}, \binits{A.M.}},
\bauthor{\bsnm{Walsh}, \binits{A.}}:
\batitle{Machine-learning structural reconstructions for accelerated point defect calculations}.
\bjtitle{npj Computational Materials}
\bvolume{10}(\bissue{1}),
\bfpage{121}
(\byear{2024})
\end{barticle}
\endbibitem

\bibitem{kolluru2022open}
\begin{barticle}
\bauthor{\bsnm{Kolluru}, \binits{A.}},
\bauthor{\bsnm{Shuaibi}, \binits{M.}},
\bauthor{\bsnm{Palizhati}, \binits{A.}},
\bauthor{\bsnm{Shoghi}, \binits{N.}},
\bauthor{\bsnm{Das}, \binits{A.}},
\bauthor{\bsnm{Wood}, \binits{B.}},
\bauthor{\bsnm{Zitnick}, \binits{C.L.}},
\bauthor{\bsnm{Kitchin}, \binits{J.R.}},
\bauthor{\bsnm{Ulissi}, \binits{Z.W.}}:
\batitle{Open challenges in developing generalizable large-scale machine-learning models for catalyst discovery}.
\bjtitle{ACS Catalysis}
\bvolume{12}(\bissue{14}),
\bfpage{8572}--\blpage{8581}
(\byear{2022})
\end{barticle}
\endbibitem

\bibitem{kim2023structure}
\begin{barticle}
\bauthor{\bsnm{Kim}, \binits{S.}},
\bauthor{\bsnm{Noh}, \binits{J.}},
\bauthor{\bsnm{Jin}, \binits{T.}},
\bauthor{\bsnm{Lee}, \binits{J.}},
\bauthor{\bsnm{Jung}, \binits{Y.}}:
\batitle{A structure translation model for crystal compounds}.
\bjtitle{npj Computational Materials}
\bvolume{9}(\bissue{1}),
\bfpage{142}
(\byear{2023})
\end{barticle}
\endbibitem

\bibitem{yoon2020differentiable}
\begin{barticle}
\bauthor{\bsnm{Yoon}, \binits{J.}},
\bauthor{\bsnm{Ulissi}, \binits{Z.W.}}:
\batitle{Differentiable optimization for the prediction of ground state structures (dogss)}.
\bjtitle{Physical Review Letters}
\bvolume{125}(\bissue{17}),
\bfpage{173001}
(\byear{2020})
\end{barticle}
\endbibitem

\bibitem{wang2024concurrent}
\begin{barticle}
\bauthor{\bsnm{Wang}, \binits{Z.}},
\bauthor{\bsnm{Wang}, \binits{X.}},
\bauthor{\bsnm{Luo}, \binits{X.}},
\bauthor{\bsnm{Gao}, \binits{P.}},
\bauthor{\bsnm{Sun}, \binits{Y.}},
\bauthor{\bsnm{Lv}, \binits{J.}},
\bauthor{\bsnm{Wang}, \binits{H.}},
\bauthor{\bsnm{Wang}, \binits{Y.}},
\bauthor{\bsnm{Ma}, \binits{Y.}}:
\batitle{Concurrent learning scheme for crystal structure prediction}.
\bjtitle{Physical Review B}
\bvolume{109}(\bissue{9}),
\bfpage{094117}
(\byear{2024})
\end{barticle}
\endbibitem

\bibitem{Omee2024}
\begin{barticle}
\bauthor{\bsnm{Omee}, \binits{S.S.}},
\bauthor{\bsnm{Wei}, \binits{L.}},
\bauthor{\bsnm{Hu}, \binits{M.}},
\bauthor{\bsnm{Hu}, \binits{J.}}:
\batitle{Crystal structure prediction using neural network potential and age-fitness pareto genetic algorithm}.
\bjtitle{Journal of Materials Informatics}
\bvolume{4}(\bissue{1}),
\bfpage{2}
(\byear{2024})
\end{barticle}
\endbibitem

\bibitem{kazeev2023sparse}
\begin{barticle}
\bauthor{\bsnm{Kazeev}, \binits{N.}},
\bauthor{\bsnm{Al-Maeeni}, \binits{A.R.}},
\bauthor{\bsnm{Romanov}, \binits{I.}},
\bauthor{\bsnm{Faleev}, \binits{M.}},
\bauthor{\bsnm{Lukin}, \binits{R.}},
\bauthor{\bsnm{Tormasov}, \binits{A.}},
\bauthor{\bsnm{Castro~Neto}, \binits{A.}},
\bauthor{\bsnm{Novoselov}, \binits{K.S.}},
\bauthor{\bsnm{Huang}, \binits{P.}},
\bauthor{\bsnm{Ustyuzhanin}, \binits{A.}}:
\batitle{Sparse representation for machine learning the properties of defects in 2d materials}.
\bjtitle{npj Computational Materials}
\bvolume{9}(\bissue{1}),
\bfpage{113}
(\byear{2023})
\end{barticle}
\endbibitem

\bibitem{mosquera2023identifying}
\begin{barticle}
\bauthor{\bsnm{Mosquera-Lois}, \binits{I.}},
\bauthor{\bsnm{Kavanagh}, \binits{S.R.}},
\bauthor{\bsnm{Walsh}, \binits{A.}},
\bauthor{\bsnm{Scanlon}, \binits{D.O.}}:
\batitle{Identifying the ground state structures of point defects in solids}.
\bjtitle{npj Computational Materials}
\bvolume{9}(\bissue{1}),
\bfpage{25}
(\byear{2023})
\end{barticle}
\endbibitem

\bibitem{huang2023unveiling}
\begin{barticle}
\bauthor{\bsnm{Huang}, \binits{P.}},
\bauthor{\bsnm{Lukin}, \binits{R.}},
\bauthor{\bsnm{Faleev}, \binits{M.}},
\bauthor{\bsnm{Kazeev}, \binits{N.}},
\bauthor{\bsnm{Al-Maeeni}, \binits{A.R.}},
\bauthor{\bsnm{Andreeva}, \binits{D.V.}},
\bauthor{\bsnm{Ustyuzhanin}, \binits{A.}},
\bauthor{\bsnm{Tormasov}, \binits{A.}},
\bauthor{\bsnm{Castro~Neto}, \binits{A.}},
\bauthor{\bsnm{Novoselov}, \binits{K.S.}}:
\batitle{Unveiling the complex structure-property correlation of defects in 2d materials based on high throughput datasets}.
\bjtitle{npj 2D Materials and Applications}
\bvolume{7}(\bissue{1}),
\bfpage{6}
(\byear{2023})
\end{barticle}
\endbibitem

\bibitem{jiang2024machine}
\begin{barticle}
\bauthor{\bsnm{Jiang}, \binits{C.}},
\bauthor{\bsnm{Marianetti}, \binits{C.A.}},
\bauthor{\bsnm{Khafizov}, \binits{M.}},
\bauthor{\bsnm{Hurley}, \binits{D.H.}}:
\batitle{Machine learning potential assisted exploration of complex defect potential energy surfaces}.
\bjtitle{npj Computational Materials}
\bvolume{10}(\bissue{1}),
\bfpage{21}
(\byear{2024})
\end{barticle}
\endbibitem

\bibitem{belsky2002new}
\begin{barticle}
\bauthor{\bsnm{Belsky}, \binits{A.}},
\bauthor{\bsnm{Hellenbrandt}, \binits{M.}},
\bauthor{\bsnm{Karen}, \binits{V.L.}},
\bauthor{\bsnm{Luksch}, \binits{P.}}:
\batitle{New developments in the inorganic crystal structure database (icsd): accessibility in support of materials research and design}.
\bjtitle{Acta Crystallographica Section B: Structural Science}
\bvolume{58}(\bissue{3}),
\bfpage{364}--\blpage{369}
(\byear{2002})
\end{barticle}
\endbibitem

\bibitem{jain2013commentary}
\begin{botherref}
\oauthor{\bsnm{Jain}, \binits{A.}},
\oauthor{\bsnm{Ong}, \binits{S.P.}},
\oauthor{\bsnm{Hautier}, \binits{G.}},
\oauthor{\bsnm{Chen}, \binits{W.}},
\oauthor{\bsnm{Richards}, \binits{W.D.}},
\oauthor{\bsnm{Dacek}, \binits{S.}},
\oauthor{\bsnm{Cholia}, \binits{S.}},
\oauthor{\bsnm{Gunter}, \binits{D.}},
\oauthor{\bsnm{Skinner}, \binits{D.}},
\oauthor{\bsnm{Ceder}, \binits{G.}}, et al.:
Commentary: The materials project: A materials genome approach to accelerating materials innovation.
APL materials
\textbf{1}(1)
(2013)
\end{botherref}
\endbibitem

\bibitem{kim2020generative}
\begin{barticle}
\bauthor{\bsnm{Kim}, \binits{S.}},
\bauthor{\bsnm{Noh}, \binits{J.}},
\bauthor{\bsnm{Gu}, \binits{G.H.}},
\bauthor{\bsnm{Aspuru-Guzik}, \binits{A.}},
\bauthor{\bsnm{Jung}, \binits{Y.}}:
\batitle{Generative adversarial networks for crystal structure prediction}.
\bjtitle{ACS central science}
\bvolume{6}(\bissue{8}),
\bfpage{1412}--\blpage{1420}
(\byear{2020})
\end{barticle}
\endbibitem

\bibitem{haastrup2018computational}
\begin{barticle}
\bauthor{\bsnm{Haastrup}, \binits{S.}},
\bauthor{\bsnm{Strange}, \binits{M.}},
\bauthor{\bsnm{Pandey}, \binits{M.}},
\bauthor{\bsnm{Deilmann}, \binits{T.}},
\bauthor{\bsnm{Schmidt}, \binits{P.S.}},
\bauthor{\bsnm{Hinsche}, \binits{N.F.}},
\bauthor{\bsnm{Gjerding}, \binits{M.N.}},
\bauthor{\bsnm{Torelli}, \binits{D.}},
\bauthor{\bsnm{Larsen}, \binits{P.M.}},
\bauthor{\bsnm{Riis-Jensen}, \binits{A.C.}}, \betal:
\batitle{The computational 2d materials database: high-throughput modeling and discovery of atomically thin crystals}.
\bjtitle{2D Materials}
\bvolume{5}(\bissue{4}),
\bfpage{042002}
(\byear{2018})
\end{barticle}
\endbibitem

\bibitem{gjerding2021recent}
\begin{barticle}
\bauthor{\bsnm{Gjerding}, \binits{M.N.}},
\bauthor{\bsnm{Taghizadeh}, \binits{A.}},
\bauthor{\bsnm{Rasmussen}, \binits{A.}},
\bauthor{\bsnm{Ali}, \binits{S.}},
\bauthor{\bsnm{Bertoldo}, \binits{F.}},
\bauthor{\bsnm{Deilmann}, \binits{T.}},
\bauthor{\bsnm{Kn{\o}sgaard}, \binits{N.R.}},
\bauthor{\bsnm{Kruse}, \binits{M.}},
\bauthor{\bsnm{Larsen}, \binits{A.H.}},
\bauthor{\bsnm{Manti}, \binits{S.}}, \betal:
\batitle{Recent progress of the computational 2d materials database (c2db)}.
\bjtitle{2D Materials}
\bvolume{8}(\bissue{4}),
\bfpage{044002}
(\byear{2021})
\end{barticle}
\endbibitem

\bibitem{lyngby2022data}
\begin{barticle}
\bauthor{\bsnm{Lyngby}, \binits{P.}},
\bauthor{\bsnm{Thygesen}, \binits{K.S.}}:
\batitle{Data-driven discovery of 2d materials by deep generative models}.
\bjtitle{npj Computational Materials}
\bvolume{8}(\bissue{1}),
\bfpage{232}
(\byear{2022})
\end{barticle}
\endbibitem

\bibitem{schutt2021equivariant}
\begin{bchapter}
\bauthor{\bsnm{Sch{\"u}tt}, \binits{K.}},
\bauthor{\bsnm{Unke}, \binits{O.}},
\bauthor{\bsnm{Gastegger}, \binits{M.}}:
\bctitle{Equivariant message passing for the prediction of tensorial properties and molecular spectra}.
In: \bbtitle{International Conference on Machine Learning},
pp. \bfpage{9377}--\blpage{9388}
(\byear{2021}).
\bcomment{PMLR}
\end{bchapter}
\endbibitem

\bibitem{xie2018crystal}
\begin{barticle}
\bauthor{\bsnm{Xie}, \binits{T.}},
\bauthor{\bsnm{Grossman}, \binits{J.C.}}:
\batitle{Crystal graph convolutional neural networks for an accurate and interpretable prediction of material properties}.
\bjtitle{Physical review letters}
\bvolume{120}(\bissue{14}),
\bfpage{145301}
(\byear{2018})
\end{barticle}
\endbibitem

\bibitem{noh2019unveiling}
\begin{barticle}
\bauthor{\bsnm{Noh}, \binits{J.}},
\bauthor{\bsnm{Kim}, \binits{S.}},
\bauthor{\bparticle{ho} \bsnm{Gu}, \binits{G.}},
\bauthor{\bsnm{Shinde}, \binits{A.}},
\bauthor{\bsnm{Zhou}, \binits{L.}},
\bauthor{\bsnm{Gregoire}, \binits{J.M.}},
\bauthor{\bsnm{Jung}, \binits{Y.}}:
\batitle{Unveiling new stable manganese based photoanode materials via theoretical high-throughput screening and experiments}.
\bjtitle{Chemical Communications}
\bvolume{55}(\bissue{89}),
\bfpage{13418}--\blpage{13421}
(\byear{2019})
\end{barticle}
\endbibitem

\bibitem{satorras2021n}
\begin{bchapter}
\bauthor{\bsnm{Satorras}, \binits{V.G.}},
\bauthor{\bsnm{Hoogeboom}, \binits{E.}},
\bauthor{\bsnm{Welling}, \binits{M.}}:
\bctitle{E (n) equivariant graph neural networks}.
In: \bbtitle{International Conference on Machine Learning},
pp. \bfpage{9323}--\blpage{9332}
(\byear{2021}).
\bcomment{PMLR}
\end{bchapter}
\endbibitem

\bibitem{zhang2023efficient}
\begin{barticle}
\bauthor{\bsnm{Zhang}, \binits{X.}},
\bauthor{\bsnm{Zhang}, \binits{O.}},
\bauthor{\bsnm{Shen}, \binits{C.}},
\bauthor{\bsnm{Qu}, \binits{W.}},
\bauthor{\bsnm{Chen}, \binits{S.}},
\bauthor{\bsnm{Cao}, \binits{H.}},
\bauthor{\bsnm{Kang}, \binits{Y.}},
\bauthor{\bsnm{Wang}, \binits{Z.}},
\bauthor{\bsnm{Wang}, \binits{E.}},
\bauthor{\bsnm{Zhang}, \binits{J.}}, \betal:
\batitle{Efficient and accurate large library ligand docking with karmadock}.
\bjtitle{Nature Computational Science}
\bvolume{3}(\bissue{9}),
\bfpage{789}--\blpage{804}
(\byear{2023})
\end{barticle}
\endbibitem

\bibitem{dong2023equivariant}
\begin{botherref}
\oauthor{\bsnm{Dong}, \binits{T.}},
\oauthor{\bsnm{Yang}, \binits{Z.}},
\oauthor{\bsnm{Zhou}, \binits{J.}},
\oauthor{\bsnm{Chen}, \binits{C.Y.-C.}}:
Equivariant flexible modeling of the protein--ligand binding pose with geometric deep learning.
Journal of Chemical Theory and Computation
(2023)
\end{botherref}
\endbibitem

\bibitem{li2022deep}
\begin{barticle}
\bauthor{\bsnm{Li}, \binits{H.}},
\bauthor{\bsnm{Wang}, \binits{Z.}},
\bauthor{\bsnm{Zou}, \binits{N.}},
\bauthor{\bsnm{Ye}, \binits{M.}},
\bauthor{\bsnm{Xu}, \binits{R.}},
\bauthor{\bsnm{Gong}, \binits{X.}},
\bauthor{\bsnm{Duan}, \binits{W.}},
\bauthor{\bsnm{Xu}, \binits{Y.}}:
\batitle{Deep-learning density functional theory hamiltonian for efficient ab initio electronic-structure calculation}.
\bjtitle{Nature Computational Science}
\bvolume{2}(\bissue{6}),
\bfpage{367}--\blpage{377}
(\byear{2022})
\end{barticle}
\endbibitem

\bibitem{yu2022uncertainty}
\begin{botherref}
\oauthor{\bsnm{Yu}, \binits{J.}},
\oauthor{\bsnm{Wang}, \binits{D.}},
\oauthor{\bsnm{Zheng}, \binits{M.}}:
Uncertainty quantification: Can we trust artificial intelligence in drug discovery?
Iscience
(2022)
\end{botherref}
\endbibitem

\bibitem{tran2018active}
\begin{barticle}
\bauthor{\bsnm{Tran}, \binits{K.}},
\bauthor{\bsnm{Ulissi}, \binits{Z.W.}}:
\batitle{Active learning across intermetallics to guide discovery of electrocatalysts for co2 reduction and h2 evolution}.
\bjtitle{Nature Catalysis}
\bvolume{1}(\bissue{9}),
\bfpage{696}--\blpage{703}
(\byear{2018})
\end{barticle}
\endbibitem

\bibitem{szymanski2023autonomous}
\begin{barticle}
\bauthor{\bsnm{Szymanski}, \binits{N.J.}},
\bauthor{\bsnm{Rendy}, \binits{B.}},
\bauthor{\bsnm{Fei}, \binits{Y.}},
\bauthor{\bsnm{Kumar}, \binits{R.E.}},
\bauthor{\bsnm{He}, \binits{T.}},
\bauthor{\bsnm{Milsted}, \binits{D.}},
\bauthor{\bsnm{McDermott}, \binits{M.J.}},
\bauthor{\bsnm{Gallant}, \binits{M.}},
\bauthor{\bsnm{Cubuk}, \binits{E.D.}},
\bauthor{\bsnm{Merchant}, \binits{A.}}, \betal:
\batitle{An autonomous laboratory for the accelerated synthesis of novel materials}.
\bjtitle{Nature}
\bvolume{624}(\bissue{7990}),
\bfpage{86}--\blpage{91}
(\byear{2023})
\end{barticle}
\endbibitem

\bibitem{li2024jmi}
\begin{barticle}
\bauthor{\bsnm{Li}, \binits{Y.}},
\bauthor{\bsnm{Wu}, \binits{Y.}},
\bauthor{\bsnm{Han}, \binits{Y.}},
\bauthor{\bsnm{Qiujie}, \binits{L.}},
\bauthor{\bsnm{Wu}, \binits{H.}},
\bauthor{\bsnm{Zhang}, \binits{X.}},
\bauthor{\bsnm{Shen}, \binits{L.}}:
\batitle{Local environment interaction-based machine learning framework for predicting molecular adsorption energy}.
\bjtitle{Journal of Materials Informatics}
\bvolume{4}(\bissue{1}),
\bfpage{4}
(\byear{2024})
\end{barticle}
\endbibitem

\bibitem{musaelian2023learning}
\begin{barticle}
\bauthor{\bsnm{Musaelian}, \binits{A.}},
\bauthor{\bsnm{Batzner}, \binits{S.}},
\bauthor{\bsnm{Johansson}, \binits{A.}},
\bauthor{\bsnm{Sun}, \binits{L.}},
\bauthor{\bsnm{Owen}, \binits{C.J.}},
\bauthor{\bsnm{Kornbluth}, \binits{M.}},
\bauthor{\bsnm{Kozinsky}, \binits{B.}}:
\batitle{Learning local equivariant representations for large-scale atomistic dynamics}.
\bjtitle{Nature Communications}
\bvolume{14}(\bissue{1}),
\bfpage{579}
(\byear{2023})
\end{barticle}
\endbibitem

\bibitem{pablo2023fast}
\begin{botherref}
\oauthor{\bsnm{Pablo-Garc{\'\i}a}, \binits{S.}},
\oauthor{\bsnm{Morandi}, \binits{S.}},
\oauthor{\bsnm{Vargas-Hern{\'a}ndez}, \binits{R.A.}},
\oauthor{\bsnm{Jorner}, \binits{K.}},
\oauthor{\bsnm{Ivkovi{\'c}}, \binits{{\v{Z}}.}},
\oauthor{\bsnm{L{\'o}pez}, \binits{N.}},
\oauthor{\bsnm{Aspuru-Guzik}, \binits{A.}}:
Fast evaluation of the adsorption energy of organic molecules on metals via graph neural networks.
Nature Computational Science,
1--10
(2023)
\end{botherref}
\endbibitem

\bibitem{gong2023general}
\begin{barticle}
\bauthor{\bsnm{Gong}, \binits{X.}},
\bauthor{\bsnm{Li}, \binits{H.}},
\bauthor{\bsnm{Zou}, \binits{N.}},
\bauthor{\bsnm{Xu}, \binits{R.}},
\bauthor{\bsnm{Duan}, \binits{W.}},
\bauthor{\bsnm{Xu}, \binits{Y.}}:
\batitle{General framework for e (3)-equivariant neural network representation of density functional theory hamiltonian}.
\bjtitle{Nature Communications}
\bvolume{14}(\bissue{1}),
\bfpage{2848}
(\byear{2023})
\end{barticle}
\endbibitem

\bibitem{zhong2023transferable}
\begin{barticle}
\bauthor{\bsnm{Zhong}, \binits{Y.}},
\bauthor{\bsnm{Yu}, \binits{H.}},
\bauthor{\bsnm{Su}, \binits{M.}},
\bauthor{\bsnm{Gong}, \binits{X.}},
\bauthor{\bsnm{Xiang}, \binits{H.}}:
\batitle{Transferable equivariant graph neural networks for the hamiltonians of molecules and solids}.
\bjtitle{npj Computational Materials}
\bvolume{9}(\bissue{1}),
\bfpage{182}
(\byear{2023})
\end{barticle}
\endbibitem

\bibitem{zhong2024universal}
\begin{botherref}
\oauthor{\bsnm{Zhong}, \binits{Y.}},
\oauthor{\bsnm{Yu}, \binits{H.}},
\oauthor{\bsnm{Yang}, \binits{J.}},
\oauthor{\bsnm{Guo}, \binits{X.}},
\oauthor{\bsnm{Xiang}, \binits{H.}},
\oauthor{\bsnm{Gong}, \binits{X.}}:
Universal machine learning kohn-sham hamiltonian for materials.
Chinese Physics Letters
(2024)
\end{botherref}
\endbibitem

\bibitem{park2020developing}
\begin{barticle}
\bauthor{\bsnm{Park}, \binits{C.W.}},
\bauthor{\bsnm{Wolverton}, \binits{C.}}:
\batitle{Developing an improved crystal graph convolutional neural network framework for accelerated materials discovery}.
\bjtitle{Physical Review Materials}
\bvolume{4}(\bissue{6}),
\bfpage{063801}
(\byear{2020})
\end{barticle}
\endbibitem

\bibitem{schutt2017schnet}
\begin{botherref}
\oauthor{\bsnm{Sch{\"u}tt}, \binits{K.}},
\oauthor{\bsnm{Kindermans}, \binits{P.-J.}},
\oauthor{\bsnm{Sauceda~Felix}, \binits{H.E.}},
\oauthor{\bsnm{Chmiela}, \binits{S.}},
\oauthor{\bsnm{Tkatchenko}, \binits{A.}},
\oauthor{\bsnm{M{\"u}ller}, \binits{K.-R.}}:
Schnet: A continuous-filter convolutional neural network for modeling quantum interactions.
Advances in neural information processing systems
\textbf{30}
(2017)
\end{botherref}
\endbibitem

\bibitem{chen2019graph}
\begin{barticle}
\bauthor{\bsnm{Chen}, \binits{C.}},
\bauthor{\bsnm{Ye}, \binits{W.}},
\bauthor{\bsnm{Zuo}, \binits{Y.}},
\bauthor{\bsnm{Zheng}, \binits{C.}},
\bauthor{\bsnm{Ong}, \binits{S.P.}}:
\batitle{Graph networks as a universal machine learning framework for molecules and crystals}.
\bjtitle{Chemistry of Materials}
\bvolume{31}(\bissue{9}),
\bfpage{3564}--\blpage{3572}
(\byear{2019})
\end{barticle}
\endbibitem

\bibitem{gasteiger_dimenet_2020}
\begin{bchapter}
\bauthor{\bsnm{Gasteiger}, \binits{J.}},
\bauthor{\bsnm{Gro{\ss}}, \binits{J.}},
\bauthor{\bsnm{G{\"u}nnemann}, \binits{S.}}:
\bctitle{Directional message passing for molecular graphs}.
In: \bbtitle{International Conference on Learning Representations (ICLR)}
(\byear{2020})
\end{bchapter}
\endbibitem

\bibitem{choudhary2021atomistic}
\begin{barticle}
\bauthor{\bsnm{Choudhary}, \binits{K.}},
\bauthor{\bsnm{DeCost}, \binits{B.}}:
\batitle{Atomistic line graph neural network for improved materials property predictions}.
\bjtitle{npj Computational Materials}
\bvolume{7}(\bissue{1}),
\bfpage{185}
(\byear{2021})
\end{barticle}
\endbibitem

\bibitem{unke2021spookynet}
\begin{barticle}
\bauthor{\bsnm{Unke}, \binits{O.T.}},
\bauthor{\bsnm{Chmiela}, \binits{S.}},
\bauthor{\bsnm{Gastegger}, \binits{M.}},
\bauthor{\bsnm{Sch{\"u}tt}, \binits{K.T.}},
\bauthor{\bsnm{Sauceda}, \binits{H.E.}},
\bauthor{\bsnm{M{\"u}ller}, \binits{K.-R.}}:
\batitle{Spookynet: Learning force fields with electronic degrees of freedom and nonlocal effects}.
\bjtitle{Nature communications}
\bvolume{12}(\bissue{1}),
\bfpage{7273}
(\byear{2021})
\end{barticle}
\endbibitem

\bibitem{batzner20223}
\begin{barticle}
\bauthor{\bsnm{Batzner}, \binits{S.}},
\bauthor{\bsnm{Musaelian}, \binits{A.}},
\bauthor{\bsnm{Sun}, \binits{L.}},
\bauthor{\bsnm{Geiger}, \binits{M.}},
\bauthor{\bsnm{Mailoa}, \binits{J.P.}},
\bauthor{\bsnm{Kornbluth}, \binits{M.}},
\bauthor{\bsnm{Molinari}, \binits{N.}},
\bauthor{\bsnm{Smidt}, \binits{T.E.}},
\bauthor{\bsnm{Kozinsky}, \binits{B.}}:
\batitle{E (3)-equivariant graph neural networks for data-efficient and accurate interatomic potentials}.
\bjtitle{Nature communications}
\bvolume{13}(\bissue{1}),
\bfpage{2453}
(\byear{2022})
\end{barticle}
\endbibitem

\bibitem{banik2023cegann}
\begin{barticle}
\bauthor{\bsnm{Banik}, \binits{S.}},
\bauthor{\bsnm{Dhabal}, \binits{D.}},
\bauthor{\bsnm{Chan}, \binits{H.}},
\bauthor{\bsnm{Manna}, \binits{S.}},
\bauthor{\bsnm{Cherukara}, \binits{M.}},
\bauthor{\bsnm{Molinero}, \binits{V.}},
\bauthor{\bsnm{Sankaranarayanan}, \binits{S.K.}}:
\batitle{Cegann: Crystal edge graph attention neural network for multiscale classification of materials environment}.
\bjtitle{npj Computational Materials}
\bvolume{9}(\bissue{1}),
\bfpage{23}
(\byear{2023})
\end{barticle}
\endbibitem

\bibitem{unke2019physnet}
\begin{barticle}
\bauthor{\bsnm{Unke}, \binits{O.T.}},
\bauthor{\bsnm{Meuwly}, \binits{M.}}:
\batitle{Physnet: A neural network for predicting energies, forces, dipole moments, and partial charges}.
\bjtitle{Journal of chemical theory and computation}
\bvolume{15}(\bissue{6}),
\bfpage{3678}--\blpage{3693}
(\byear{2019})
\end{barticle}
\endbibitem

\bibitem{zhang2022interpretable}
\begin{barticle}
\bauthor{\bsnm{Zhang}, \binits{X.}},
\bauthor{\bsnm{Zhou}, \binits{J.}},
\bauthor{\bsnm{Lu}, \binits{J.}},
\bauthor{\bsnm{Shen}, \binits{L.}}:
\batitle{Interpretable learning of voltage for electrode design of multivalent metal-ion batteries}.
\bjtitle{npj Computational Materials}
\bvolume{8}(\bissue{1}),
\bfpage{175}
(\byear{2022})
\end{barticle}
\endbibitem

\bibitem{omee2022scalable}
\begin{botherref}
\oauthor{\bsnm{Omee}, \binits{S.S.}},
\oauthor{\bsnm{Louis}, \binits{S.-Y.}},
\oauthor{\bsnm{Fu}, \binits{N.}},
\oauthor{\bsnm{Wei}, \binits{L.}},
\oauthor{\bsnm{Dey}, \binits{S.}},
\oauthor{\bsnm{Dong}, \binits{R.}},
\oauthor{\bsnm{Li}, \binits{Q.}},
\oauthor{\bsnm{Hu}, \binits{J.}}:
Scalable deeper graph neural networks for high-performance materials property prediction.
Patterns
(2022)
\end{botherref}
\endbibitem

\bibitem{haghighatlari2022newtonnet}
\begin{barticle}
\bauthor{\bsnm{Haghighatlari}, \binits{M.}},
\bauthor{\bsnm{Li}, \binits{J.}},
\bauthor{\bsnm{Guan}, \binits{X.}},
\bauthor{\bsnm{Zhang}, \binits{O.}},
\bauthor{\bsnm{Das}, \binits{A.}},
\bauthor{\bsnm{Stein}, \binits{C.J.}},
\bauthor{\bsnm{Heidar-Zadeh}, \binits{F.}},
\bauthor{\bsnm{Liu}, \binits{M.}},
\bauthor{\bsnm{Head-Gordon}, \binits{M.}},
\bauthor{\bsnm{Bertels}, \binits{L.}}, \betal:
\batitle{Newtonnet: A newtonian message passing network for deep learning of interatomic potentials and forces}.
\bjtitle{Digital Discovery}
\bvolume{1}(\bissue{3}),
\bfpage{333}--\blpage{343}
(\byear{2022})
\end{barticle}
\endbibitem

\bibitem{han2024survey}
\begin{botherref}
\oauthor{\bsnm{Han}, \binits{J.}},
\oauthor{\bsnm{Cen}, \binits{J.}},
\oauthor{\bsnm{Wu}, \binits{L.}},
\oauthor{\bsnm{Li}, \binits{Z.}},
\oauthor{\bsnm{Kong}, \binits{X.}},
\oauthor{\bsnm{Jiao}, \binits{R.}},
\oauthor{\bsnm{Yu}, \binits{Z.}},
\oauthor{\bsnm{Xu}, \binits{T.}},
\oauthor{\bsnm{Wu}, \binits{F.}},
\oauthor{\bsnm{Wang}, \binits{Z.}}, et al.:
A survey of geometric graph neural networks: Data structures, models and applications.
arXiv preprint arXiv:2403.00485
(2024)
\end{botherref}
\endbibitem

\bibitem{yang2022learning}
\begin{barticle}
\bauthor{\bsnm{Yang}, \binits{Z.}},
\bauthor{\bsnm{Zhong}, \binits{W.}},
\bauthor{\bsnm{Lv}, \binits{Q.}},
\bauthor{\bsnm{Chen}, \binits{C.Y.-C.}}:
\batitle{Learning size-adaptive molecular substructures for explainable drug--drug interaction prediction by substructure-aware graph neural network}.
\bjtitle{Chemical Science}
\bvolume{13}(\bissue{29}),
\bfpage{8693}--\blpage{8703}
(\byear{2022})
\end{barticle}
\endbibitem

\bibitem{yang2022mgraphdta}
\begin{barticle}
\bauthor{\bsnm{Yang}, \binits{Z.}},
\bauthor{\bsnm{Zhong}, \binits{W.}},
\bauthor{\bsnm{Zhao}, \binits{L.}},
\bauthor{\bsnm{Chen}, \binits{C.Y.-C.}}:
\batitle{Mgraphdta: deep multiscale graph neural network for explainable drug--target binding affinity prediction}.
\bjtitle{Chemical science}
\bvolume{13}(\bissue{3}),
\bfpage{816}--\blpage{833}
(\byear{2022})
\end{barticle}
\endbibitem

\bibitem{vaswani2017attention}
\begin{botherref}
\oauthor{\bsnm{Vaswani}, \binits{A.}},
\oauthor{\bsnm{Shazeer}, \binits{N.}},
\oauthor{\bsnm{Parmar}, \binits{N.}},
\oauthor{\bsnm{Uszkoreit}, \binits{J.}},
\oauthor{\bsnm{Jones}, \binits{L.}},
\oauthor{\bsnm{Gomez}, \binits{A.N.}},
\oauthor{\bsnm{Kaiser}, \binits{{\L}.}},
\oauthor{\bsnm{Polosukhin}, \binits{I.}}:
Attention is all you need.
Advances in neural information processing systems
\textbf{30}
(2017)
\end{botherref}
\endbibitem

\bibitem{liberti2014euclidean}
\begin{barticle}
\bauthor{\bsnm{Liberti}, \binits{L.}},
\bauthor{\bsnm{Lavor}, \binits{C.}},
\bauthor{\bsnm{Maculan}, \binits{N.}},
\bauthor{\bsnm{Mucherino}, \binits{A.}}:
\batitle{Euclidean distance geometry and applications}.
\bjtitle{SIAM review}
\bvolume{56}(\bissue{1}),
\bfpage{3}--\blpage{69}
(\byear{2014})
\end{barticle}
\endbibitem

\bibitem{lu2022tankbind}
\begin{barticle}
\bauthor{\bsnm{Lu}, \binits{W.}},
\bauthor{\bsnm{Wu}, \binits{Q.}},
\bauthor{\bsnm{Zhang}, \binits{J.}},
\bauthor{\bsnm{Rao}, \binits{J.}},
\bauthor{\bsnm{Li}, \binits{C.}},
\bauthor{\bsnm{Zheng}, \binits{S.}}:
\batitle{Tankbind: Trigonometry-aware neural networks for drug-protein binding structure prediction}.
\bjtitle{Advances in neural information processing systems}
\bvolume{35},
\bfpage{7236}--\blpage{7249}
(\byear{2022})
\end{barticle}
\endbibitem

\bibitem{masters2023deep}
\begin{barticle}
\bauthor{\bsnm{Masters}, \binits{M.R.}},
\bauthor{\bsnm{Mahmoud}, \binits{A.H.}},
\bauthor{\bsnm{Wei}, \binits{Y.}},
\bauthor{\bsnm{Lill}, \binits{M.A.}}:
\batitle{Deep learning model for efficient protein--ligand docking with implicit side-chain flexibility}.
\bjtitle{Journal of Chemical Information and Modeling}
\bvolume{63}(\bissue{6}),
\bfpage{1695}--\blpage{1707}
(\byear{2023})
\end{barticle}
\endbibitem

\bibitem{gawlikowski2023survey}
\begin{barticle}
\bauthor{\bsnm{Gawlikowski}, \binits{J.}},
\bauthor{\bsnm{Tassi}, \binits{C.R.N.}},
\bauthor{\bsnm{Ali}, \binits{M.}},
\bauthor{\bsnm{Lee}, \binits{J.}},
\bauthor{\bsnm{Humt}, \binits{M.}},
\bauthor{\bsnm{Feng}, \binits{J.}},
\bauthor{\bsnm{Kruspe}, \binits{A.}},
\bauthor{\bsnm{Triebel}, \binits{R.}},
\bauthor{\bsnm{Jung}, \binits{P.}},
\bauthor{\bsnm{Roscher}, \binits{R.}}, \betal:
\batitle{A survey of uncertainty in deep neural networks}.
\bjtitle{Artificial Intelligence Review}
\bvolume{56}(\bissue{Suppl 1}),
\bfpage{1513}--\blpage{1589}
(\byear{2023})
\end{barticle}
\endbibitem

\bibitem{luo2023calibrated}
\begin{botherref}
\oauthor{\bsnm{Luo}, \binits{Y.}},
\oauthor{\bsnm{Liu}, \binits{Y.}},
\oauthor{\bsnm{Peng}, \binits{J.}}:
Calibrated geometric deep learning improves kinase--drug binding predictions.
Nature Machine Intelligence,
1--12
(2023)
\end{botherref}
\endbibitem

\bibitem{kresse1996efficient}
\begin{barticle}
\bauthor{\bsnm{Kresse}, \binits{G.}},
\bauthor{\bsnm{Furthm{\"u}ller}, \binits{J.}}:
\batitle{Efficient iterative schemes for ab initio total-energy calculations using a plane-wave basis set}.
\bjtitle{Physical review B}
\bvolume{54}(\bissue{16}),
\bfpage{11169}
(\byear{1996})
\end{barticle}
\endbibitem

\bibitem{uhrin2021workflows}
\begin{barticle}
\bauthor{\bsnm{Uhrin}, \binits{M.}},
\bauthor{\bsnm{Huber}, \binits{S.P.}},
\bauthor{\bsnm{Yu}, \binits{J.}},
\bauthor{\bsnm{Marzari}, \binits{N.}},
\bauthor{\bsnm{Pizzi}, \binits{G.}}:
\batitle{Workflows in aiida: Engineering a high-throughput, event-based engine for robust and modular computational workflows}.
\bjtitle{Computational Materials Science}
\bvolume{187},
\bfpage{110086}
(\byear{2021})
\end{barticle}
\endbibitem

\bibitem{sungwon_kim_2023_8081655}
\begin{botherref}
\oauthor{\bsnm{Kim}, \binits{S.}}:
{A Structure Translation Model for Crystal Compounds Release for manuscript acceptance}
(2023).
\doiurl{10.5281/zenodo.8081655}
\end{botherref}
\endbibitem

\bibitem{Chen2022_19470599}
\begin{botherref}
\oauthor{\bsnm{Chen}, \binits{C.}},
\oauthor{\bsnm{Ong}, \binits{S.P.}}:
{MPF.2021.2.8}
(2022).
\doiurl{10.6084/m9.figshare.19470599.v3}
\end{botherref}
\endbibitem

\bibitem{yang_2024_13160937}
\begin{botherref}
\oauthor{\bsnm{Yang}, \binits{Z.}},
\oauthor{\bsnm{Zhao}, \binits{Y.-M.}},
\oauthor{\bsnm{Wang}, \binits{X.}},
\oauthor{\bsnm{Liu}, \binits{X.}},
\oauthor{\bsnm{Zhang}, \binits{X.}},
\oauthor{\bsnm{Li}, \binits{Y.}},
\oauthor{\bsnm{Lv}, \binits{Q.}},
\oauthor{\bsnm{Chen}, \binits{C.Y.-C.}},
\oauthor{\bsnm{Shen}, \binits{L.}}:
{Source Code for ``Scalable Crystal Structure Relaxation Using an Iteration-Free Deep Generative Model with Uncertainty Quantification"}
(2024).
\doiurl{10.5281/zenodo.13160937}
\end{botherref}
\endbibitem

\end{thebibliography}


\end{document}